\documentclass[%
 reprint,
 amsmath,amssymb,
 aps,
]{revtex4-2}

\usepackage[]{svg}
\svgpath{{../}}
\usepackage{graphicx}
\usepackage{pifont}
\usepackage{overpic}

\usepackage{float}
\usepackage{dcolumn}
\usepackage{bm}
\usepackage{hyperref}

\hypersetup{
    colorlinks=true,
    linkcolor=blue,
    filecolor=blue,      
    urlcolor=blue,
    citecolor=blue
    }

\usepackage{xcolor}
\usepackage{subfiles} 
\usepackage{ulem}

\renewcommand{\emph}[1]{\textit{#1}}
\begin{document}

\preprint{APS/123-QED}

\title{Steady Granular Flow in a Rotating Drum:\\ 
Universal description of stress, velocity and packing fraction profiles covering grain shape effects from convex to very concave.}

\author{Weiyi Wang \textsuperscript{1}}
\email{weiyi.wang@umontpellier.fr}

\author{Jonathan Barés \textsuperscript{1}}%
\email{jonathan.bares@umontpellier.fr}

\author{Mathieu Renouf \textsuperscript{1}}%
\email{mathieu.renouf@umontpellier.fr}

\author{Emilien Azéma \textsuperscript{1,2,}}%
\email{emilien.azema@umontpellier.fr}
 
\affiliation{
 \textsuperscript{1} LMGC, Université de Montpellier, CNRS, Montpellier, France
}%

\affiliation{
 \textsuperscript{2} Institut Universitaire de France (IUF), Paris, France\\
}%

\date{\today}

\begin{abstract}
The flow behavior of granular matter is significantly influenced by the shape of constituent particles. This effect is particularly pronounced for very concave particles, which exhibit unique flow characteristics such as higher porosity and sharper phase transitions between jamming and unjamming states. Despite the richness and ubiquitousness of these systems, our understanding of their intricate flow behavior and the local mechanisms driving these behaviors remains incomplete. In this work, we investigate the effect of particle shape, ranging from spherical to highly concave, on steady flows in a rotating drum - a system that facilitates a continuous phase transition from a jamming state at greater depths to an unjamming state at shallower regions. We develop an analytical model to elucidate granular behavior within the rotating drum:
(i) Firstly, by decomposing the shear stress, we reconcile the discrepancy between simulation data and theoretical predictions, establishing a relationship with the angle of repose.
(ii)Secondly, we extend the generalized Bagnold scaling , coupled with a non-local fluidity relation based on packing fraction, providing a framework for a correlation between shear stress, shear rate, and packing fraction. Additionally, we introduce a characteristic length to quantify the influence of particle shape and drum speed. 
This analytical model offers explicit functional forms for physical quantity profiles, which are validated experimentally in a thin rotating drum and numerically in a two-dimensional rotating drum. Our results demonstrate that this model accurately describes the change of velocity due to the phase transition of granular flow within a rotating drum. Moreover, for different shapes of particle and drum speeds, the characteristic length captures the interplay between shear stress, shear rate, and the variation of packing fraction.
\end{abstract}

\maketitle

\section{\label{sec:INTRODUCTION}State of the art}
\subsection{Introduction: flow within a rotating drum as a textbook case}

Granular matter is ubiquitous in both natural and industrial environments, with more or less complexly shaped particles. Systems composed of these materials are commonly observed and studied both at the macroscopic and microscopic scales. 
One of their main characteristics is their ability to exhibit both solid-like (jammed) and fluid-like (unjammed) behavior \cite{Jaeger1996,GDRMiDi2004,Forterre2008}. Still in this latter state, they are far from evolving as a well-behaved Newtonian fluid. Indeed, in contrast to the deep-rooted constitutive relationships like the Navier-Stokes equations that aptly describe liquid flow, no unique analytical model describes the evolution of any flowing granular system. In other words, the rheological behavior in granular flows diverges from that of traditional fluids, representing an active area of intense investigation in several fields for decades \cite{GDRMiDi2004,Forterre2008,Goyon2008,Kamrin2012,Tapia2019}.

Maybe one of the most complicated situations for this kind of flow is when at least one surface of the system is not constrained and is free to move as the one observed at the top of a rotating drum. In this latter case, there exist distinct flow modes that can simultaneously cohabit within the system. 
More specifically, in the so-called ``rolling regime'' (\textit{i.e.}, a flow governed by inertial effects with a well-defined dynamics \cite{Ding2001,Bonamy2002,Jain2002,GDRMiDi2004}), we observe a \emph{solid-like} behavior close to the walls and deep in bulk, where the particle motion is slow and considered as static. In contrast, in the upper part, a liquid-like behavior develops, characterized by layers of grains with different degrees of inertia. In these layers, the flow evolves from a \emph{quasi-static} regime just above the solid-like zone to \emph{inertial} in the layers close to the free surface with an almost exponential increase in grain velocity. In this last layer, the flow is characterized by a nearly linear increase of grain velocity \cite{Rajchenbach2000,Roux2002,Orpe2001,Bonamy2002,Jain2002,GDRMiDi2004}. Finally, a collisional behavior is observed at the free surface (the liquid-like regime being just below) where particles mainly interact via binary shocks \cite{Bonamy2002,Jain2002}. The transitions between each of these regimes, within a single system, are still a matter of debate \cite{Bonamy2002,GDRMiDi2004,Renouf2005,Orpe2007}: At what depth do they occur? How thick they are? How do these thicknesses vary as a function of the drum speed and size? What about the properties of the grains? 

This set of seemingly simple questions reveals a complex and still open problem: There is no well-posed constitutive law accounting for \textit{(i)} the phase-flow transition and \textit{(ii)} local effects allowing to predict the stress and velocity fields. This is still true in one of the simplest and most common flow geometry: \emph{the rotating drum}. To address these issues, and more generally to provide a more scientific description of multi-directional granular flows in rotating drums, several paths may be explored.

\subsection{\label{sec:INTRODUCTION_scal}Scaling law approach}

First, scaling laws have been formulated to relate some macroscopic observables to microscopic features of the flows. These laws are most of the time established heuristically or empirically. For instance, in the Govender's comprehensive review \cite{Govender2016}, it is shown that the mean flow velocity $\langle v\rangle$ scales with the liquid-like flow thickness $h$ as $\langle v\rangle \propto h^m$, where $m$ varies from $1$ to $3$ depending on the different observations \cite{Rajchenbach2000,GDRMiDi2004,Jop2005,Jop2006,Felix2007}. The dynamic angle of repose $\langle \theta \rangle$ is also found to vary as a power-law of angular speed $\Omega$ with exponent between $1$ and $2.6$ depending on the system properties \cite{Rajchenbach1990,Khosropour2000,Felix2007}. 
 
Richer scaling laws have also been proposed, combining the Froude number with other system parameters to describe the flow rates \cite{Taberlet2006,Florent2012,Orozco2020} or the dynamic angle of repose \cite{Orozco2020}. 

Still, despite these long-lasting efforts to raise a state equation ruling these systems, no consensus has emerged yet. Building a single equation that encompasses all the different flow scenarios is actually extremely challenging given the multitude of system parameters at play and the inherent flow diversity.

\subsection{\label{sec:INTRODUCTION_cons}The $\mu(I)$-rheology and its limitation in multi-phase flows}

To analytically describe the evolution of a granular flow, a second way actively pursued consisted of building constitutive laws that better capture the main features of the flow rheology. After decades of efforts, compiling experimental and numerical data on flows in multiple configurations, a consensus has emerged in the form of the so-called ``$\mu(I)$-rheology'' \cite{GDRMiDi2004}. This law empirically stated that the apparent macroscopic friction coefficient, $\mu=\tau/P$, and the packing fraction, $\phi$, both depend on the so-called ``Inertial number'', $I$, \cite{DaCruz2005, Hatano2007, Rycorft2009}, with $\tau$ being the shear stress and $P$ the confining pressure. This latter state parameter, $I$, is defined as the ratio of the shear time, $\dot \gamma^{-1}$, imposed by the flow rate, $\dot \gamma$, over the particle relaxation time, $(\rho d/P)^{1/2}$, for a particle of density, $\rho$, and equivalent diameter, $d$ \cite{GDRMiDi2004}. It is shown that the shear stress is proportional to the pressure, through the effective friction coefficient as $\tau = \mu(I) P$, and the volume fraction, $\phi$, is a function of $I$ as $\phi=\phi(I)$. In general, a linear dependence is obtained for small values of $I$ both for $\mu(I)$ and $\phi(I)$ \cite{DaCruz2005}. More advanced functions have also been proposed to consider larger values of $I$ \cite{Jop2005, Jop2006, Forterre2008} or values depending on the geometry \cite{Renouf2005, Hatano2007, Bonamy2009}. 

It is remarkable that this model agrees with Bagnold's theory. Indeed, assuming that the momentum transfer between particles in adjacent layers results from instantaneous binary collisions during flow, \cite{Campbell1990,Forterre2018,Tapia2019} proposed:
\begin{equation}
     \begin{matrix} 
     \tau= f_1(\phi)  \rho d^2 \dot \gamma^2 & (a) \\
     P= f_2(\phi)  \rho d^2 \dot \gamma^2 & (b)
    \end{matrix} ,
\label{Eq:initial_Bagn}
\end{equation}
where $f_1$ and $f_2$ are functions depending only on $\phi$ such that: $f_1(\phi)=\mu(I(\phi))/I^2(\phi)$ and $f_2=I^{-2}(\phi)$.
We note that Eq.\eqref{Eq:initial_Bagn} has been recovered using dimensional analysis \cite{Campbell1990} and shown to be valid for all shear rates \cite{Lois2005}.

The strength of the $\mu(I)$ model relies on the fact that when combined with a continuous approach (for the stress field) it correctly predicts the velocity fields in various homogeneous flow geometries \cite{Forterre2008, Rycorft2009}.
Nevertheless, despite this success, a number of limitations still remain \cite{Ertas2002, GDRMiDi2004, Forterre2008}. 
Indeed, for instance, this model manages to characterize the rheology of the flow in the bulk material but fails near the walls, or for finite-size flows where boundary effects dominate \cite{Miller2013, Rognon2014}. 
In the inclined plane flow, for example, the bottom and free surfaces are close and the angle at which the flowing layer comes to rest is determined by the layer thickness independently from the $\mu(I)$ prediction \cite{Silbert2001, GDRMiDi2004}. Similarly, this model cannot accurately describe the transitions both, from the quasi-static to the static regime and from the inertial to the collisional one \cite{Forterre2008}.

\begin{figure}
\centering
  \includegraphics[width=0.5\textwidth]{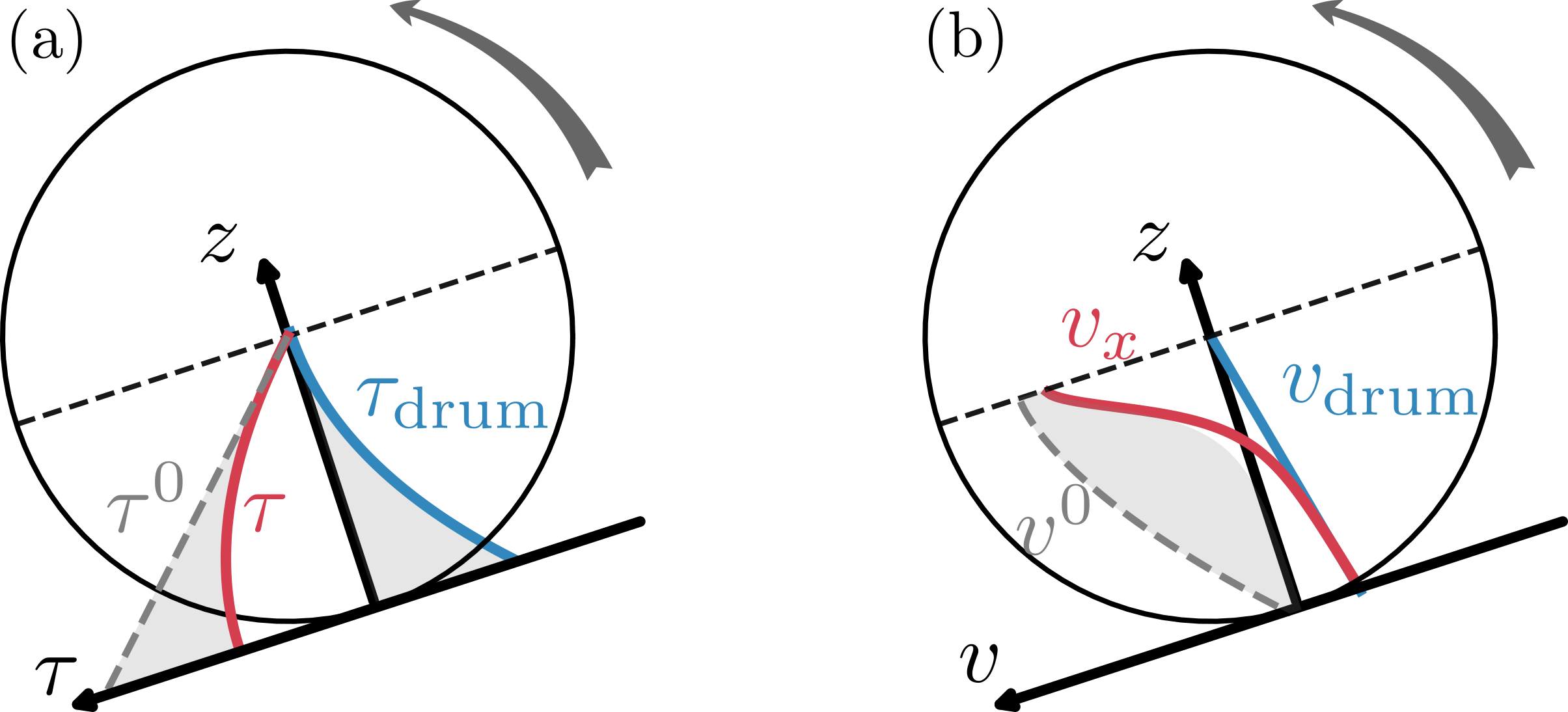}
\caption{
(a): Typical shear stress profile $\tau(z)$ as a function of depth measured in 
drum geometry (red line). The gray zone represents the deviation of $\tau(z)$ from $\tau_0$ (gray dashed line) the shear stress given by Eq.\eqref{Eq:momentum_balance}.
The blue line represent the shear stress $\tau_{drum}$ assumed inherent of the drum geometry (See discussion in Sec.\ref{Sec_Model_stress}).
(b): Typical velocity profile $v_x(z)$ as a function of depth measured in 
drum geometry (red line). The gray zone represents the deviation of $v_x(z)$ from $v_0$ (gray dashed line) the typical Bagnold-like profile for a free surface flow as in inclined plane geometry.
The blue line represent the velocity profile imposed by the drum $v_{\text{drum}}(z)$ (See also Eq.\eqref{Sec_Model_stress}).}
\label{fig_1}
\end{figure}

This latter point is particularly critical when dealing with the flow in a rotating drum since these different regimes coexist at the same time. In this flowing geometry, the collinearity between the deviatoric shear stress and strain tensor is not verified which avoids the definition of a general tensorial form of the constitutive law as in a unidirectional flow \cite{Forterre2008, Cortet2009}. Still, the scalar form of the $\mu(I)$ relationship is verified in the whole drum \cite{Renouf2005, Lin2020, Povall2021}. 
However, as evidenced in a lot of studies \cite{Renouf2005, Florent2012, Lin2020, Povall2021}, this does not mean that the stress and velocity fields can be predicted by a continuum approach with  a Bagnold-like profile as classically done for free-surface flow. 

Another approach is still possible. By assuming a permanent flow, momentum balance equations along and normal to the flow direction yield a linear stress distribution across the depth as \cite{GDRMiDi2004}:
\begin{equation}
    \left\{\begin{matrix} 
      \tau(z)=\rho g (H-z) \sin \langle \theta \rangle & (a) \\ 
      P(z)= \rho g (H-z) \cos \langle \theta \rangle, & (b)
    \end{matrix}\right.
\label{Eq:momentum_balance}
\end{equation}
where $g$ is the gravitational acceleration and $H$ is the height of the free surface along the $z$-direction (see Fig. \ref{fig_1}). 
These relations are verified for flows down to an inclined plane \cite{GDRMiDi2004, Forterre2008} and even for heap flows \cite{GDRMiDi2004}. However Eq.\eqref{Eq:momentum_balance}a fails in the case of flow within a rotating drum \cite{Renouf2005, Orpe2007}, while Eq.\eqref{Eq:momentum_balance}b continues to be valid.

In Fig. \ref{fig_1}(a) we present a sketch of the shear stress profile measured in the stationary rolling regime (full red curve) and its prediction from Eq.\eqref{Eq:momentum_balance}(a) (grey dashed line). This latter, although acceptable in the first inertial layers, differs from the measurement deep in the quasi-static and static layers. In these layers, it tends to become independent of depth. It is often argued that the discrepancy between measurement and predictions comes from the fact that the flow is not perfectly homogeneous. As a major consequence, the effective coefficient of friction, $\mu$, also varies in depth, and thus cannot be directly related to free surface angle, $\langle \theta \rangle$ \cite{Renouf2005, Orpe2007, Forterre2008}. 
Then, a constitutive law for the shear stress within a rotating drum is still missing.

Moreover, the streamwise velocity profile sketched in Fig.\ref{fig_1}(b), shows a logarithmic-like decrease as it extends further from the inertial-flow zone deep into the static bulk zone. It moves away from the Bagnolian-like profile deduced from the $\mu(I)$-rheology \cite{Orpe2007, Forterre2008}. Consequently, the $\mu(I)$ law does not apply to the drum case, requiring a \emph{non-local} extension of the Bagnold scaling.

\subsection{\label{sec:INTRODUCTION_non_local}Non-local rheology and high order fluidity parameter}
\subsubsection{Generalities}
For some flow geometries, the previous observations emphasize the need to consider the non-local nature of the granular flow. To put it simply, what occurs at a given point in the flow is inherently influenced by what diffuses from its immediate vicinity. To implement this in a coherent manner, a characteristic length scale, $l_m$, related to the grain diameter has been introduced. For each grain, it defines a range of interactions that is virtually smaller or larger than its actual geometry. Thus, a coefficient $\alpha$ is introduced such that $l_m=\alpha d$. This is assumed to account for the occurrence of collective motion of some clusters at different scales \cite{Prandtl1925, Orpe2001, Ertas2002, GDRMiDi2004, DeGiuli2015, Saitoh2020} and location within the flow. Then, following \cite{GDRMiDi2004} Bagnold Eq.\eqref{Eq:initial_Bagn} is modified as:

\begin{equation}
\begin{split}
\tau(z) &=  l_m^2(z) \rho \dot \gamma^2(z),
\end{split}
\label{eq:mixing_Bagnold_l}
\end{equation}

In the case of the rotating drum, it has been shown that this coherent length scale, $l_m$ goes to zero at the free surface. Then it progressively extends as it approaches the transition into the static phase leading to a divergence close to the walls. This behavior drastically differs from other homogeneous-like flows where $l_m$ remains most of the time relatively constant into the bulk \cite{GDRMiDi2004}. The challenge therefore lies in finding the ``right'' physical model to best describe how this coherent length scale originates and evolves as a function of the flow conditions.

As explained in the extensive reviews presented by Bouzid et al. \cite{Bouzid2015} and Kamrin \cite{Kamrin2019}, a fruitful way to overcome this challenge has consisted of building an extra parameter describing the local state of the system. This is known as the concept of \emph{``fluidity''}. It relies on the introduction of an order parameter, $f$, describing the dynamical transition from solid-like to fluid-like behavior. This latter can be understood as a field of phase-change process that quantifies the degree of fluidity displayed by a given region of a granular system. This parameter, $f$, satisfies a partial Ginzburg-Landau-like phenomenological equation based on a given diffusion process of the form \cite{Aranson2002,Bouzid2013,Kamrin2015}:
\begin{equation}
\underbrace{T \frac{D f}{D t}}_{\text{evolution, material derivative}} =
\underbrace{l^2\nabla^2 f}_{\text{diffusive term}}+
\underbrace{\mathcal{I}(f)}_{\text{source term}},
\label{eq:GinzburgLandau}
\end{equation}
where $T$ and $l$ are characteristic time and length scales, respectively. $\mathcal{I}(f)$ is a ``source term'' designed to switch the stability of the phase $f$ from solid-like to liquid-like behaviors. Under this framework, Eq.\eqref{eq:mixing_Bagnold_l} is often rewritten in a more general form as $f = \mathcal{F}(\tau,\dot \gamma,P,I...)$. Then, the main issue consists of finding the control parameters and the source function, $\mathcal{I}(f)$, that fit with the system dynamics. 

One of the earliest approaches is the so-called \emph{``Partially Fluidized Theory''} (PFT), where $f$ is assumed to vary from $1$, for solid-like behavior, to $0$ for fluid-like behavior. The characteristic length scale, $l$, is taken as the mean grain diameter \cite{Aranson2001, Aranson2002, Volfson2003}. The PFT manages to go further than the traditional $\mu(I)$ model and captures the solid-to-fluid phase transition, particularly concerning the initial and stopping heights in inclined plane \cite{Aranson2001,Aranson2002} or some specific flow patterns \cite{Aranson2006,Malloggi2006}. 

Then, the \emph{``Non Local Granular Fluidity''} (NGF) approach has emerged as a powerful tool for quantitatively predicting various non-local phenomena in various loading geometries \cite{Kamrin2012, Kamrin2015, Liu2018}. The NGF can be seen as an extension of the work of Goyon et al. \cite{Goyon2008} and Bocquet et al. \cite{Bocquet2009} where the fluidity is assumed to be the inverse of the viscosity: $f= \dot \gamma / \mu$. Also, the source term, $\mathcal{I}(f)$, is chosen to vanish when $f$ tends to a local fluidity $f_{loc}$ that depends on $\mu$. The characteristic length scale, $l$, represents long-range interactions diverging as the flow approaches the solid phase and going to $0$ in the fluidized zone. More recently, Zhang and Kamrin \cite{Zhang2017} have formulated a microscopic interpretation that has then been verified experimentally \cite{fazelpour2022_sm} and found to be equivalent to the macroscopic form (under given boundary conditions) by Poon and colleagues \cite{poon2023_pre}. It turns out that the fluidity originates in the spatial fluctuations of the individual particle velocities.
Nevertheless, the NGF is based on an explicit formulation of the $\mu(I)$ law which makes its applicability quite tricky in the drum geometry where stress information is missing. Moreover, both $\mu(z)$ and $I(z)$ are non-uniformly distributed in the whole drum. 

Finally, it is worth mentioning that other strategies have been imagined, for instance, by assuming that the fluidity is only defined from the inertial number (\textit{i.e.}, $f=I$ together with an expansion of the $\mu(I)$ law) \cite{Bouzid2013,Bouzid2015}.

\subsubsection{Discussion: the case of the rotating drum}
The development of non-local models for the flow in the rotating drum geometry (in particular to predict the stress and velocity fields) is still largely an open question. Nevertheless, it is fair to mention that some promising successes have been achieved (at least for the ``rolling'' regime) with the PFT approach. For instance, Horpe and Khakhar showed that the PFT (as developed initially by Arranson and Tsimring \cite{Aranson2001,Aranson2002}) reproduces the velocity field within a rotating drum \cite{Orpe2007}, although they do not perfectly fit the numerical/experimental data. Nonetheless, their model relies on four \textit{ad-hoc} adjustable parameters, together with a preliminary fit of the shear stress $\tau$, in the absence of a predefined stress model. We also note that their velocity model does not follow a Bagnolian profile.

In the same time as Aranson and Tsimring \cite{Aranson2002}, Bonamy and Mills \cite{Bonamy2003} developed their own non-local model. In this latter, the flowing granular medium is assumed to behave like a network of granular chains embedded in a  Newtonian viscous-like fluid. 
By decomposing the total stress as a linear combination of Coulombic friction stresses, viscous stresses and stresses due to immersed chains, they propose a simple equation based on solely two parameters (the shear rate and a characteristic length linked to the particle diameter). Their equation fits the data well in the liquid flow region, but does not describe the region near the free surface and fails to reproduce the transition to the solid region, particularly at high rotation speeds. At the same time, Bonamy revisited the PFT approach, and modified the original model by forcing the viscosity to follow a Bagnold's law \cite{bonamy_thesis}. He obtained a model that fits the experimental data much better than the earlier version, but the justification for maintaining a Bagnolian approach is missing.

\subsection{\label{sec:INTRODUCTION_shape}Grain shape effects}
It is worth noting that most of the results discussed in the previous sections have been obtained for model granular materials, \textit{i.e.} materials composed of discs (2D) or spheres (3D). Whether they are minimalist like the $\mu(I)$ model, or more sophisticated as the \emph{``fluidity''} concept, most of the models are all able to explain complex collective phenomena, but they mainly ignore the fact that the complexity can arise from the materials itself. The intrinsic properties of the grains composing the system can dramatically change its rheology \cite{Cleary2008, CEGEO2012, Athanassiadis2014, Gravish2018, Rakotonirina2019, Marschall2020, cox2016_epl, zhao2019_gm}.

For example, in the quasi-static limit (\textit{i.e.} $I<<1$), the shear strength $\mu$ increases with the grain angularity. But for larger grain angularities it may saturate towards a maximum value \cite{Emilien2012, Emilien2013_2} or even decrease toward values close to that of an assembly of discs, depending on the contact friction value \cite{Theechalit2020}.
Numerous systematic studies have also highlighted non-linear variations in the solid fraction with grain elongation \cite{Cleary2008, CEGEO2012} or grain shape non-convexity \cite{CEGEO2012, Emilien2013, Athanassiadis2014, Rakotonirina2019, Conzelmann2022}. In the case of homogeneous flows, it appears that the $\mu(I)$-rheology is still well observed, at least for convex \cite{Emilien2012_2} or slightly concave \cite{Emilien2013,Han2021} grains. This leads to a simple translation of the trends. 

However, for the same thickness, the avalanche initiation angle is greater for non-spherical grain packing \cite{Emilien2012_2,Han2021}, demonstrating that the shape of the grains modifies the local properties. In silo-like geometries, faster discharges were observed for elongated particles than for rounded particles \cite{Tang2016}. In contrast, grain discharges slow down with increasingly angular grains \cite{Goldberg2018}. This again reveals non-local effects where elongated grains manifest nematic ordering  \cite{Hidalgo2009,Emilien2012_3}, facilitating their ejection, whereas increasingly pronounced arching effects occur with angular grains induced by face-to-face contact \cite{Emilien2012}. In drum geometry two distinct flow regimes were observed depending on rotation speed and grain shape \cite{Hohner2014}: (\textit{i}) at high rotation speeds irregularly shaped grains led to a higher dynamic angle of repose caused by the granular packing expansion, (\textit{ii}) at low rotation speeds, on the contrary, the angle is more related to the packing fraction. For elongated/flat particles, differences in the velocity profile measured in the active layer were reported as the aspect ratio increases \cite{Ma2018}. Similarly, non-spherical particles showed less axial dispersion than glass beads \cite{Dube2013}.

Among all the possible grain shapes, concave grains are of particular interest. 
Indeed, highly concave particles demonstrate steep transitions between different dynamic states \cite{Marschall2020}. They can even maintain stability and loading without any external confinement through their interlocking capability and long-range correlation \cite{zhao2016_gm, bares2017_epj, Dierichs2016, Murphy2017}. At the same time, and contradictorily, they exhibit very low packing fraction \cite{Athanassiadis2014, Conzelmann2022}.  For instance, the column stability of concave U-shaped particles shows that the decrease in particle packing fraction offsets the increase in entanglement with concavity, these two trends conspire to generate a maximum of resistance to separation in collections of nonconvex particles of intermediate aspect ratio \cite{Gravish2018}. Very recently, it has also been reported that sheared hexapods develop a secondary flow profile that is completely opposite to that of convex grains in the same geometry \cite{Mohammadi2022}. This makes the systems composed of these particles an ideal system for exploring the fundamental principles governing phase transitions from grain shape properties. 

The task is very challenging. In addition to theoretical aspects aimed at generalizing rheological and non-local models, systematic studies need to be developed to continuously assess the effect of grain shape change (from convex to highly concave) on rheological properties. From an experimental point of view, a fundamental challenge is to design a sufficient number of particles while systematically controlling their shapes and mechanical properties. In addition, accounting for particle shape in numerical simulations using discrete element methods presents a number of technical hurdles that are both geometrically and computationally complex. For example, one of these challenges relates to the detection of contacts and the calculation of forces between particles of arbitrary shape, particularly concave ones, which can have a prohibitive number of contacts.

\subsection{\label{sec:INTRODUCTION_outline}Objectives and outline of the paper}

In this work, we investigate the effect of grain shape non-convexity on the intrinsic rheology of a ``rolling'' steady-state flow within a rotating drum.
To this end, we design 2D numerical studies and 3D experiments in which the grain shape is systematically varied from circular (and spherical, respectively) to cross-shaped particles with very thin arms. The drum speed is also systematically varied to explore different rolling regimes. Ultimately, we aim to develop a non-local model for predicting the stress and velocity fields. This model has to be sufficiently general to capture the effects that grain shape can have on the different flow regimes (depth and thickness) observed in the drum geometry. As mentioned above, the geometry of the rotating drum is chosen to serve as an essential experimental device to study the influence of highly concave particles, in particular in continuous phase transitions. 

In Section \ref{sec:MODEL}, we develop a stress distribution model for the drum and extend the Bagnold scaling using a non-local fluidity parameter model, which enables us to describe the velocity profile of particle flow in the drum. In particular, we show that the fluidity parameter is fully determined by the variation of the packing fraction across the drum. Consequently, we develop an equation for the flow velocity that is fully determined by well-defined parameters linked to the thickness of the different layers in the flow of the drum. In Section \ref{sec:2DSIMUS}, we test the results of 2D simulations that validate the stress, velocity, and fluidity profiles obtained with our model. We also investigate the effect of the particle shape on the granular flow behavior. In Section \ref{sec:EXP}, we present experimental results that also validate the velocity profile and investigate the influence of particle shape on 3D granular flows. Finally, we conclude in Section \ref{sec:CONCLUSION} and draw some perspectives.

\section{\label{sec:MODEL}ANALYTICAL MODEL}
\subsection{Notations and main assumptions}
\begin{figure}
\centering
\includegraphics[width=0.3\textwidth]{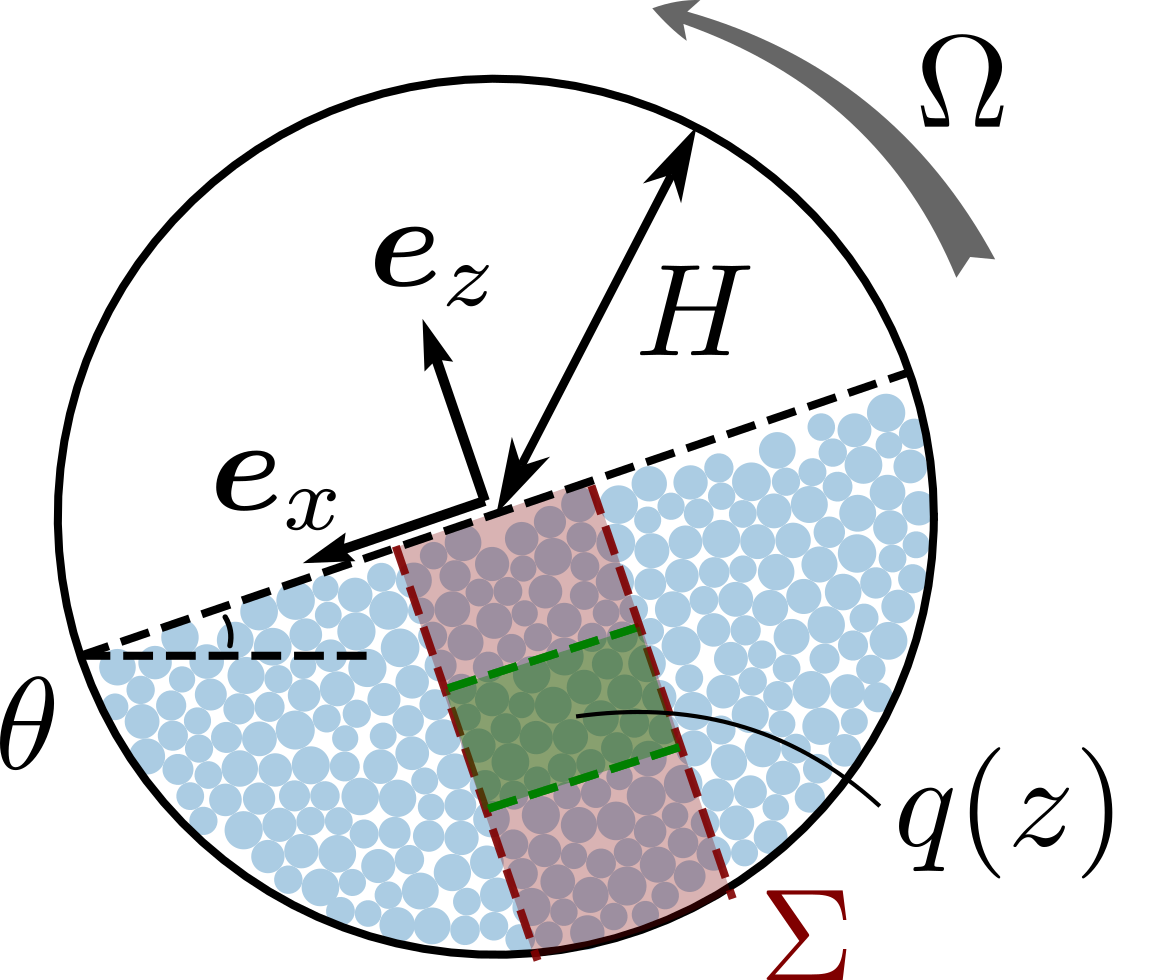}
\caption{
Illustration of the set-up. A granular system is flowing in a half-filled rotating drum of radius, $H$, and depth, $W$. The unit vector $\pmb{e}_x$ oriented with an angle $\theta$ from the horizontal is parallel to the free surface and $\pmb{e}_z$ is perpendicular to it. $\Sigma$ (in red) is a narrow area in which the flow is considered invariant along $x$ in the steady state regime. $q(z)$ represents the value of any system observable, averaged in space and time in the green area.)}
\label{fig_2_new}
\end{figure}

In the rotating drum shown in fig.\ref{fig_2_new}, we define the reference frame $\mathcal{R}_0=(\pmb{e}_x,\pmb{e}_z)$, where $\pmb{e}_x$ (resp. $\pmb{e}_z$) is parallel (resp. perpendicular) to the free surface. This latter is inclined of an angle $\theta$ with the horizon. In the reference frame, the flow can be considered as quasi-homogeneous at the center of the drum, within a thin elementary slice $\Sigma$ (red area in fig.\ref{fig_2_new}), parallel to $\pmb{e}_z$ and located around $x=0$. The slice can be divided into elementary horizontal layers (green area in fig.\ref{fig_2_new}) stacked on top of each other and parallel to the flow. The value of a given continuum quantity $q(z)$ (packing fraction, velocity, stress, \textit{etc.}) at depth $z$ is then defined as the average of the corresponding quantity given at the grain scale for all the particles in all the frames of the sequence whose center of mass is inside the layer. We assume that the flow is sufficiently steady and quasi-homogeneous across the flowing layer within the slice $\Sigma$, so that: 
\begin{equation}
\left\{\begin{matrix} 
      \frac{\partial q(z)}{\partial t} \simeq 0 \\ 
      \frac{\partial q(z)}{\partial x} \simeq 0 
    \end{matrix}\right.
\label{Eq:assumption}
\end{equation}

We also assume the packing fraction to be a global variable, $\phi$, in the granular flow. We define it as the ratio of the volume of the particles to the volume of the packing, $\phi =V_{\text{grain}}/V_{\text{packing}}$. 

Finally, the drum depth, $W$, is assumed to be wide enough with respect to the apparent diameter of the grains, $d$, to prevent side effects induced by lateral walls.

\subsection{Stress profile}
\label{Sec_Model_stress}
Starting from the Cauchy momentum equation that describes the non-relativistic momentum transport in any continuum so in granular flow, we state that \cite{acheson1991_bk}:
\begin{equation}
\underbrace{\frac{\partial \boldsymbol{v}}{\partial t}+(\boldsymbol{v} \cdot \nabla) \boldsymbol{v}}_{\text{material derivative, } D\pmb{v}/Dt}
=
\frac{1}{\rho} \underbrace{(\nabla \cdot \boldsymbol{\tau} - \nabla P)}_{\nabla \cdot \pmb{\sigma}} +\mathbf{g}
\label{eq:cauchy}
\end{equation}
where $\boldsymbol{v}$ is the velocity vector field of the flow (depending on time, $t$, and space, $(x,z)$), $\rho$ is the material density, $\mathbf{g}$ is the gravitational acceleration, and $\pmb{\sigma}$ is the stress tensor that we can decompose into $P$ the hydrostatic pressure and $\boldsymbol{\tau}$ the deviatoric stress tensor.

When studying the drum geometry, it is quite classical to decompose the velocity $\boldsymbol{v}$ into two distinct sources using the superposition principle \cite{GDRMiDi2004, Renouf2005, Orpe2007, Cortet2009}: one linear component following the same profile as the rotating drum, $\boldsymbol{v}_{\text{d}}$, (blue curve in fig.\ref{fig_1}(b)), and another, non-linear, representing the velocity relative to the drum, $\boldsymbol{v}_r$.
This gives:
\begin{equation}
\boldsymbol{v}=\boldsymbol{v}_r+\boldsymbol{v}_{\text{d}}
\label{eq:relative_v}
\end{equation}

According to the first principle of special relativity, stating that all physical laws take their simplest form in an inertial frame and that there exist multiple inertial frames interrelated by uniform translation \cite{lorentz1952_bk}, we postulate that the same holds true for the stress tensor leading to the following decomposition:
\begin{equation}
\boldsymbol{\sigma}=\boldsymbol{\sigma}_r+\boldsymbol{\sigma}_{\text{d}}
\label{eq:relative_sigma}
\end{equation}
where $\boldsymbol{\sigma}_r$, is the stress tensor relative to the drum, and $\boldsymbol{\sigma}_{\text{d}}$ the stresses resulting from the rotation induced by the drum. However, while in Eq.\eqref{eq:relative_v} at least $\boldsymbol{v}_{\text{d}}$ is well defined, in Eq.\eqref{eq:relative_sigma} neither $\boldsymbol{\sigma}_{\text{d}}$ nor $\boldsymbol{\sigma}_r$ are known {\it a priori}. 

As a first approximation, in the drum reference frame ($\mathcal{R}_0$), the flow can be considered similar to the flow on a heap (assuming no lateral wall effects \cite{Pouliquen2002_2,Jop2005}). 
This analogy can be invoked thanks to the so-called self-similarity of the velocity profiles in these two configurations \cite{Pouliquen2002_2, GDRMiDi2004, Hung2016}.
Thus, according to our main assumptions (see eq.\eqref{Eq:assumption}), plugging $\boldsymbol{\sigma}_r$ and $\boldsymbol{v}_r$ within Eq.\ref{eq:cauchy} in the reference frame $\mathcal{R}_0$ leads naturally to the following equations (see also Eq.\ref{Eq:momentum_balance} in the State of the Art section):
\begin{equation}
    \left\{\begin{matrix} 
      \tau_r(z)=\rho g (H-z) \sin \langle \theta \rangle & (a) \\ 
      P_r(z)= \rho g (H-z) \cos \langle \theta \rangle, & (b)
    \end{matrix}\right.
\label{Eq:tauRPR}
\end{equation}
Where $\tau_r(z)$ and $P_r(z)$ are the average shear and normal stresses components of $\boldsymbol{\sigma_r}$, respectively. It should be noted that Eq.\ref{Eq:tauRPR} is always verified in heap flow geometry from the free surface deep into the static bulk \cite{Ray2021}. 
Thus, since the repose angle $\langle \theta \rangle$ remains constant, on average, within $\mathcal{R}_0$, the relative effective coefficient of friction $\mu_r$ remains constant too and writes as:
\begin{equation}
\mu_r=\frac{\tau_r(z)}{P_r(z)}=\tan \langle \theta \rangle.
\label{eq:friction_r}
\end{equation}

Moreover, it is also reported in the literature that Eq.\ref{Eq:tauRPR}(b) fits correctly the normal stress distribution in the rotating drum geometry \cite{Renouf2005}. Thus, it can be stated that $P(z) = P_r(z)$, where $P(z)$ is the hydrostatic component of $\boldsymbol{\sigma}$. This implies that $\boldsymbol{\sigma}_{\text{d}}$ carries only the shear stress $\tau_{\text{d}}$.

Therefore, we can define a ``basal'' friction coefficient $\mu_d$. This latter is transmitted only by the drum and results from wall effects that propagate through the material \cite{Andreotti2013_bk}. Accordingly, it can be written as:
\begin{equation}
\tau_{\text{d}}(z)= -\mu_d P(z)
\label{eq:tau_annular}
\end{equation}
The negative sign comes from the fact that the orientation of $\tau_{\text{drum}}(z)$ must correspond with the outer layer drag speed. 
Identifying the basal friction law is a complex task and has led to different strategies depending on the studied geometry \cite{Pouliquen2002, Andreotti2013_bk, breard2020_jgr}.

In our case, we formulate a set of statements. (\textit{i}) First, by construction,  
$\mu + \mu_d = \mu_r = \tan \langle \theta \rangle$,  where $\mu=\tau/P$, with $\tau$ the shear stress extracted from $\boldsymbol{\sigma}$ 
Since it is known that $\mu=\mu(z)$ evolves linearly with $z$ in the drum geometry \cite{Lin2020} (See also Appendix \ref{A_profil_mu}), then $\mu_d=\mu_d(z)$. More accurately $\mu_d$ is linear with $z$. (\textit{ii}) Second, $\mu_d$ is assumed to vanish at the free surface for a sufficiently large system, meaning that: $\mu_d|_{z=H}=0$. (\textit{iii}) Third, near the bottom wall $\mu_d|_{z=0}=\mu_w$, with $\mu_w\in]0,\tan \langle \theta \rangle[$ an effective coefficient of friction between the flowing layer and the bottom. All these conditions lead to the following:
\begin{equation}
\mu_d(z)=\mu_w\frac{H-z}{H}.
\label{eq:friction_assumption}
\end{equation}
Consequently, plugging this expression of $\mu_d$ in Eq.\ref{eq:tau_annular} gives:
\begin{equation}
\tau_{\text{d}}(z)= -\mu_w P(z) \frac{H-z}{H} 
\label{eq:tau_drum2}
\end{equation}
Interestingly, Eq.\ref{eq:tau_drum2} is reminiscent of the shear stress profiles induced by the walls in vertical-chute flows \cite{Nedderman1992, GDRMiDi2004, Artoni2007, Artoni2011}. This implies a non-trivial analogy where, at least for stresses, the flow within a rotating drum can be viewed as an intricated combination of heap flow and pipe flow. Finally, noting that the basal conditions near the wall are necessarily the same for $\mu$ and $\mu_d$ (\textit{i.e.}, $\mu|_{z=0} = \mu_d|_{z=0}$), we get that $\mu_w= \frac{1}{2} \tan \langle \theta \rangle$. Thus, plugging this last expression into Eq.\ref{eq:tau_drum2}, together with Eq.\ref{Eq:tauRPR}b (recalling that $P(z)=P_r(z)$), we get an explicit formulation of $\tau_{\text{drum}}$ of the form:
\begin{equation}
\begin{split}
\tau_{\text{d}}(z)&= -\rho g\frac{(H-z)^2}{2H}\text{cos} \langle \theta \rangle 
\end{split}
\label{eq:tau_drum}
\end{equation}
A sketch of the evolution of $\tau_{\text{drum}}(z)$ is shown in blue in Fig.\ref{fig_1}a.
Then, combining Eq.\ref{eq:tau_drum} and Eq.\ref{Eq:tauRPR} together with Eq.\ref{eq:relative_sigma}, the shear stress component $\tau$ within a rotating drum writes as:
\begin{equation}
\begin{split}
\tau(z) &= \rho g \frac{H^2-z^2}{2H}\text{sin}\langle \theta \rangle
\end{split}
\label{eq:tau}
\end{equation}

Finally, we define the granular flow density at coordinate $z$, $\rho(z)$, using the packing fraction of the assembly, $\phi(z)$, and the grain density $\rho_0$: $\rho(z) = \rho_0 \phi(z)$. We are then able to scale the shear and normal stresses to state their theoretical expression within a rotating drum as follows:
\begin{equation}
      \frac{\tau(z)}{\rho_0 gd}= \phi(z) \frac{H^2-z^2}{2Hd}\sin  \langle \theta \rangle
\label{Eq_tau2}
\end{equation}
\begin{equation}
      \frac{P(z)}{\rho_0 gd}= \phi(z) \frac{H-z}{d}\cos \langle \theta \rangle.
\label{eq:Pressure2}
\end{equation}

\subsection{Velocity profile: a non-local model based on the packing fraction}
\label{Sec_Model_velocity_fluidity}
In this section, based on recent work using the PFT approach, we develop a non-local velocity model that is ({\it i}) compatible with Bagnold's scaling law and ({\it ii}) accounts for the effect of the grain shape.  

\subsubsection{Preliminary step: equivalence between fluidity and packing fraction profiles}
\label{subsub_fluidity_Sec}
\begin{figure}
\centering
\includegraphics[width=0.95\linewidth]{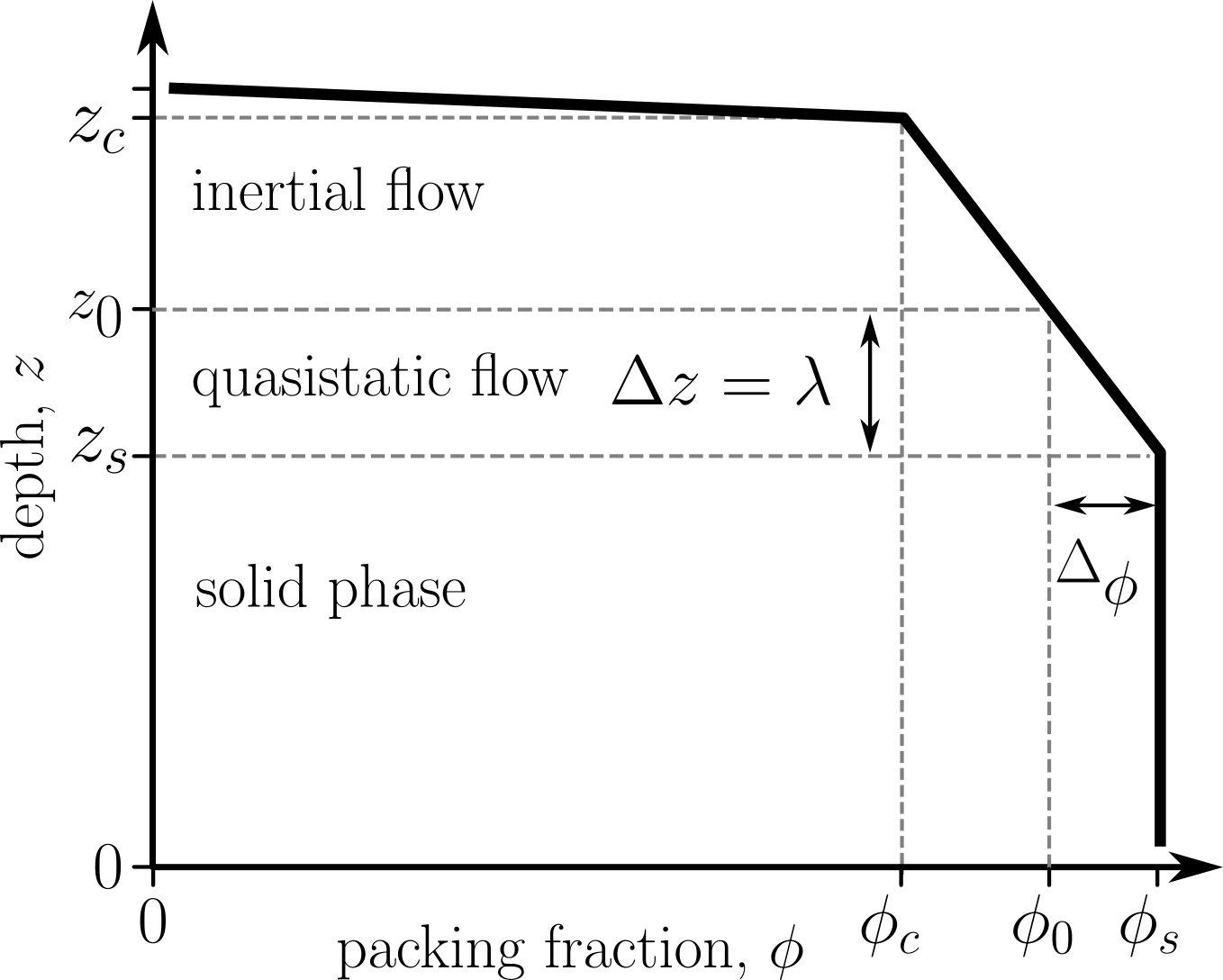}
\caption{Illustration of the packing fraction profile, $\phi(z)$, as expressed by Eq.\ref{eq:phi_generic}, where $z_s$, $z_0$ and $z_c$ are the depths at the transition from solid, to quasi-static and finally collisional states respectively. 
$\lambda=z_0-z_s$ is the thickness of the quasi-static flow zone.}
\label{fig_3}
\end{figure}

We start from the Ginzburg-Landau-like phenomenological equation \ref{eq:GinzburgLandau} presented in Sec.\ref{eq:GinzburgLandau}. Following Aranson and Tsimring \cite{Aranson2002}, the source term can be chosen as $\mathcal{I}(f) = f(1-f)(f-\delta)$, where $\delta$ is a function living between $0$ and $1$ defined as: $\delta=(\mu-\mu_{dyn})/(\mu_{sta}-\mu_{dyn})$. It is built from $\mu_{sta}$ and $\mu_{dyn}$
the dynamic and static friction coefficient respectively, \textit{i.e.}, the friction coefficient of a given granular material measured at small and high inertial number, respectively. This choice is motivated by the simple fact that the source terms must have extrema both at $f=0$ (fluid-like) and $f=1$ (solid-like). The parameter $\delta$ is introduced to control the range in which both, static and dynamic phases coexist. In practice, $\delta$ is set to $0.5$. 
Under the hypothesis of a steady and fully developed flow given in Eq. \ref{Eq:assumption}, and the definition of the source term given just above, an explicit expression for $f$ can be derived in the form \cite{Aranson2002}:
\begin{equation}
f(z)=\frac{1}{2}\{1-\tanh(\frac{z-z_0}{\sqrt{8}l})\},
\label{eq:orig_f}
\end{equation}
where $z_0$ is such that $f(z_0)=\frac{1}{2}$. This corresponds approximately, with the depth at the transition between the quasi-static and the inertial flow regimes (see Fig.\ref{fig_3}). Note that, unlike Aranson and Tsimring \cite{Aranson2002}, we do not assume that the characteristic length $l$ is related to the mean diameter $d$ of the grains. 
On the contrary, we assume that $l$ is a critical length depending on the shape of the grains. Therefore this length scale is considered as a free parameter in our approach.

It is also well documented in the literature that the evolution of the packing fraction $\phi(z)$ within a rotating drum follows a relatively simple trend as a function of $z$, as depicted in Fig. \ref{fig_3} \cite{Pouliquen2002_2, Cortet2009}. 
In this figure, $\phi$ is constant and equals to $\phi_s$ in the solid phase of the drum. Then, it decreases linearly with $z$ in the liquid phase from $\phi_s$ at $z=z_s$ (\textit{i.e.} at the transition from the static to the liquid regime) to $\phi_c$ at $z=z_c$ (\textit{i.e.} at the transition from the liquid to the collisional regime). Finally, it diverges at the surface, for $z>z_c$.
Since $z_0$ is assumed to be the depth at the transition between the quasi-static and inertial regimes, we define $\lambda=z_0-z_s$ the thickness of the quasi-static flow zone (see Fig.\ref{fig_3}). From these definitions, it is easy to obtain the following expression that gives the evolution of the packing fraction in depth:
\begin{equation}
\frac{\phi_0 - \phi(z)}{\phi_s - \phi_0} = \frac{z - z_0}{\lambda},
\label{eq:phi_generic}
\end{equation}
where $\phi_0=\phi(z_0)$. Now, multiplying the left- and right-hand side of Eq.\ref{eq:phi_generic} by $\xi = \lambda/(\sqrt{8}l)$ and plugging it into Eq.\ref{eq:orig_f} we obtain a definition of the fluidity, $f$, where the dependence on $z$ appears only through $\phi(z)$. This permits to define a functional $F$ depending only on $\phi$:
\begin{equation}
F(\phi) = \frac{1}{2}\{1-\tanh(\xi \frac{\phi_0-\phi}{\phi_s-\phi_0})\} \equiv f(\phi(z))
\label{eq:Fphi}
\end{equation}
Thus, we see that the fluidity parameter can only be described in terms of the packing fraction $\phi$ considering the functional, $F$, through the global parameter $\xi$. This latter allows to linking of two, {\it a priori}, unknowns parameters that both depend on the grain properties (\textit{i.e.} shapes, sizes...): the thickness of the quasi-static flow, $\lambda$, and the characteristic length parameter, $l$. It should be noted that, considering that $F(\phi)\rightarrow 1$ when $\phi \rightarrow \phi_s$, together with a quick look at the form of the $\tanh(\xi)$ function, this suggests that $\xi$ can be chosen to be greater than at least $2$, but in the same time cannot tend to the infinity since $\lambda<z_c-z_s$. Thus, $\xi$ is a fitting parameter that has to be determined numerically by adjusting Eq.\ref{eq:Fphi} on Eq.\ref{eq:orig_f}, and, contrary to $\lambda$ and $l$, it is not expected to depend on the grain properties. We will note it as an ``universal'' parameter.

\subsubsection{Velocity profile built from the fluidity}
Now we aim to generalize the Bagnold scaling by coupling it with the fluidity function.
Settled on the works mentioned in the State of the Art Section (Sec.\ref{sec:INTRODUCTION}), two remarks can be made. 

On the one hand, it appears that in its general Bagnolian form, the shear stress is related to the flow rate through a function $f_1(\phi)$ that depends on the packing fraction (see Eq.\ref{Eq:initial_Bagn}a). Actually, several studies have evidenced that the function $f_1$ exhibits a divergence as the packing fraction $\phi$ approaches its maximum value $\phi_s$. More precisely, there is evidence that it is necessary for $f_1(\phi)$ to scale with $(\phi_s-\phi)^{-2}$ in the vicinity of $\phi_s$ \cite{Forterre2018, Tapia2019}.
Thus, incorporating the concept of fluidity into Eq.\ref{Eq:initial_Bagn}a together with the aforementioned conditions on $f_1(\phi)$ a natural expression of $f_1(\phi)$ is:
\begin{equation}
\begin{split}
f_1(\phi) = \frac{k^2}{(1-F(\phi))^2},
\label{eq:f_1_fluidity}
\end{split}
\end{equation}
where $k$ is a dimensionless constant that does not depend, {\it a priori}, on the grain properties that are naturally captured by $\phi$. $k$ can be considered as a second ``universal'' parameter. Note that, the square form of $k$ in this equation will be justified.  

On the other hand, and as already discussed, several authors have proposed to modify the Bagnolian equation by introducing a characteristic length scaling, through a parameter $\alpha$, with the particle diameter: $l_m(z)=\alpha(z) d$. This scaling depends, {\it a priori}, on the flow depth $z$ (see Eq.\ref{eq:mixing_Bagnold_l}). Thus, by assuming that this description relies essentially on the same physical mechanisms as those of Eq.\ref{Eq:initial_Bagn}a, we can assume that $\alpha^2(z)= f_1(\phi(z))$, and thus we get:
\begin{equation}
\begin{split}
\alpha(z) = \frac{k}{1-F(\phi(z))} = \frac{k}{1-f(z)},
\label{eq:mixlength1}
\end{split}
\end{equation}

Therefore, we can generalize the Bagnold scaling from Eq.\ref{eq:mixing_Bagnold_l} and Eq.\ref{eq:mixlength1}, which elucidates the intricated interplay between shear stress, shear rate, and packing fraction in the form:
\begin{equation}
(1-f(z))^2\tau(z) =\rho (kd)^2 {(\frac{\partial v_x(z)}{\partial z})}^2
\label{eq:gBagnold}
\end{equation}

Finally, putting both, the expression of the shear stress (Eq.\ref{Eq_tau2}) and the expression of the fluidity (Eq.\ref{eq:orig_f}) into Eq.\ref{eq:gBagnold}, and reminding that $\lambda = z_0-z_s = \xi \sqrt{8}l$, we obtain the following expression for the derivative of the velocity profile:

\begin{align}
\begin{split}
\frac{\partial v_x(z)}{\partial z}  
= &\frac{1}{2kd}
\sqrt{g\frac{H^2-z^2}{2H}\sin\langle\theta\rangle}
\Big(1+ \\ &\tanh(\xi\frac{z - z_s}{\lambda} - \xi)\Big)
-\Omega
\end{split}
\label{eq:shear_rate}
\end{align}

Thus, we are able to state a theoretical expression for the velocity field of the flow within a rotating drum. The above equation is based on a total of four parameters that must be adjusted. Two of them -- $k$ and $\xi$ -- are presumed to be independent of the grain properties. On the contrary, the two others -- $z_s$ and $\lambda$ -- depend on the grain shape and describe the different thicknesses/depths of the flow zones. For instance, $z_s$ can be easily determined from the packing fraction profile. To determin the three other parameters, we can rely on the relationship between $f(z)$ and $F(\phi)$ evidenced by Eq.\ref{eq:Fphi} since we see that Eq. \ref{eq:orig_f} can be simply rewritten as a function of $\xi$ and $\lambda$ only.

\section{\label{sec:2DSIMUS}NUMERICAL VALIDATION}

\subsection{Discrete element modeling, system parameters, and steady state}

Two-dimensional simulations were carried out using the Contact Dynamics (CD) method \cite{Moreau1999}. The CD method is a Discrete-Element Method (DEM) in which small-scale effects are considered into non-regularized contact laws together with a non-smooth formulation of the particle dynamics. In other words, contrary to the strategy adopted by the Molecular Dynamic approaches (also called soft-DEM) \cite{Schlick1996}, no elastic repulsive potential nor smoothing of the Coulomb friction law is used to determine contact forces. The unknown variables are the particle velocities and contact forces simultaneously found using a (parallelized \cite{Renouf2004_NSCD}) iterative algorithm based on a nonlinear Gauss-Seidel scheme. Finally, the equations of motion are integrated by an implicit time-stepping scheme. This method is numerically unconditionally stable and particularly well adapted for simulations of a large assembly of frictional particles. This is especially true for particles of complex shapes with a potentially important number of contacts. It has been extensively employed for the simulations of granular materials in two and three dimensions 
For a detailed description of the CD method, see \cite{Moreau1999}. For our simulations, we used the simulation platform LMGC90 developed in our lab \cite{duboisLMGC90,SRAMD21}.

\begin{figure}[t!]
  \centering
  \begin{minipage}[t]{1\linewidth}
  \includegraphics[width=1\textwidth,clip]{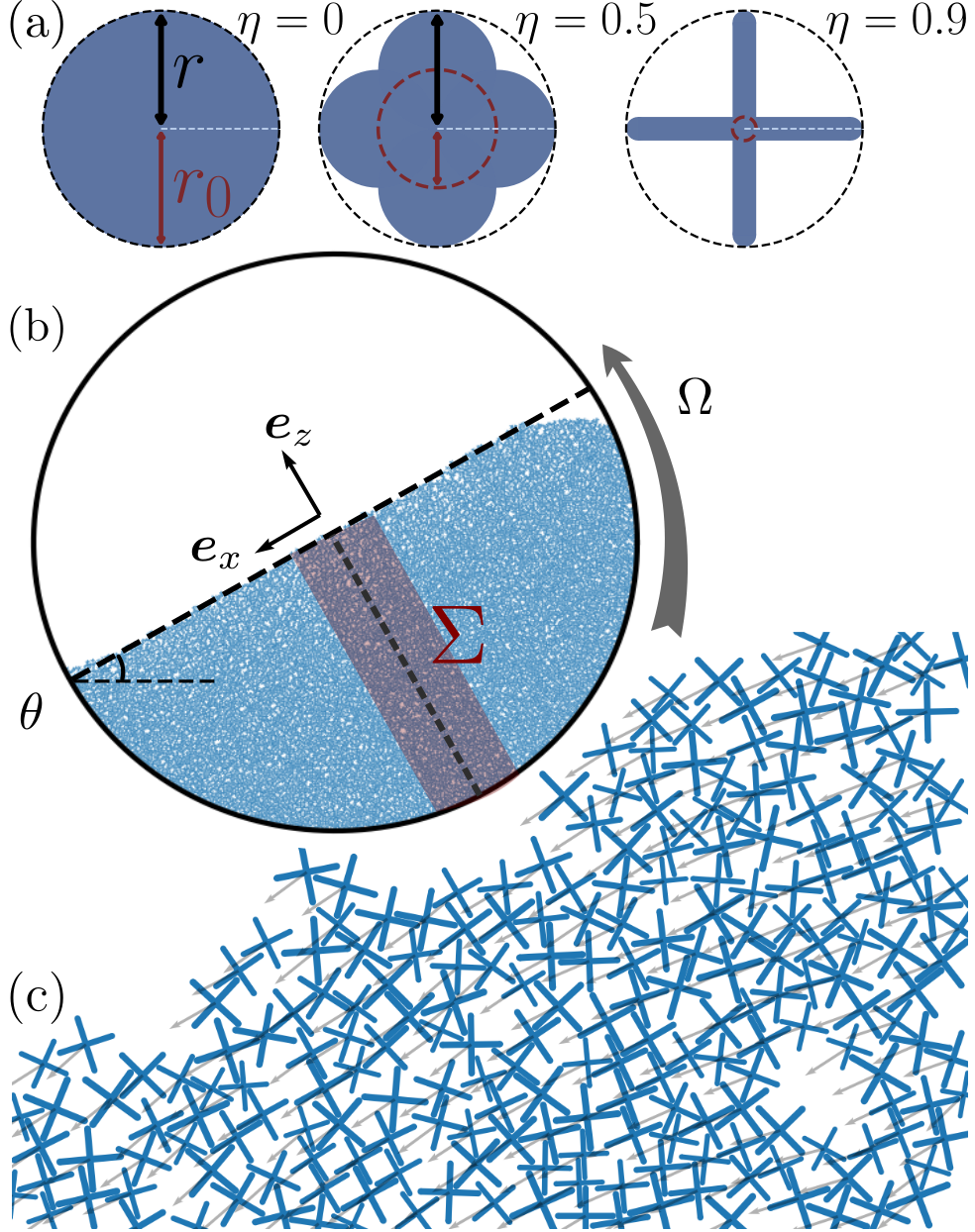}
  \end{minipage}
  \caption{(a) Example of 2D star-shaped particles with varying concavities ($\eta$).
   (b) Snapshot of the rotating drum in the permanent rolling flow regime. The red zone shows the area $\Sigma$ in which the averages are calculated. c) Zoom within the rotating sample for $\eta=0.9$. The arrows show the velocity field.}
  \label{fig:grans_2d}
\end{figure}

In this study, we consider cross-shaped particles with rounded-cap ends on each branch extremities as illustrated in Fig.\ref{fig:grans_2d}(a). Such shapes can be easily described using a ``concavity'' parameter \cite{CEGEO2012} defined as: 
\begin{equation}
    \eta=\frac{r-r_0}{r}
    \label{Eq_concavity_parameter}
\end{equation}
where $r$ is the radius of the circumscribed disk and $2 r_0$ is the branch thickness. 
For technical reasons, we consider two ways of modeling particles. In the cases where $\eta \leq 0.5$, the grains are modeled as four overlapped disks of radius $r_0$ whose centers lay at the corners of a square of edge $r_0 \sqrt(2)$. In the cases where $\eta>0.5$, the grains are built with two rectangles of length $L=2(r-r_0)$ and four disks of radius $r_0$ laying at the ends of the rectangles.
We note that the contacts between two cross-shaped grains can be reduced to a combination of contacts between disks for $\eta \leq 0.5$. In contrast, for $\eta>0.5$ three situations may arise: cap-to-cap, cap-to-line, and line-to-line contacts. Cap-to-cap and cap-to-line contacts are considered as one contact point (\textit{i.e.}, disk-disk or disk-polygon contacts, respectively). 
In the framework of the CD method, it is common to represent line-to-line contacts as two contact points. This is what is done here by considering two cap-to-line contacts. The implementation of line-to-line contacts in the framework of the CD method is described in detail in Ref. \cite{Emilien2012}. 
In the following, the concavity parameter $\eta$ is varied from $0$ (disk) to $0.9$ in steps of $0.1$.

$N_p$ randomly oriented grains of radius $r$ are first laid under the action of the gravity within a drum of radius $150r$ (see Fig.\ref{fig:grans_2d}(b)). The number, $N_p$, of grains ranges from $9 104$ for $\eta=0$, to $20707$ for $\eta=0.9$ so that the drum is half-filled whatever $\eta$. Note that a small particle size distribution is introduced around the mean radius $\bar r$ to avoid crystallization. Typically, the smallest particles have a radius of $r=1.6$~mm while the largest have $r=2.4$~mm. The friction between grains is set to $0.2$ while the friction between the grains and the drum is set to $0.9$. This prevents slipping at the boundaries. Finally, the particle mass is kept constant by adjusting the material density, $\rho_0$, for each $\eta$.  

\begin{figure}[t!]
  \begin{minipage}[t]{0.9\linewidth}
  \begin{overpic}[width=1\textwidth]{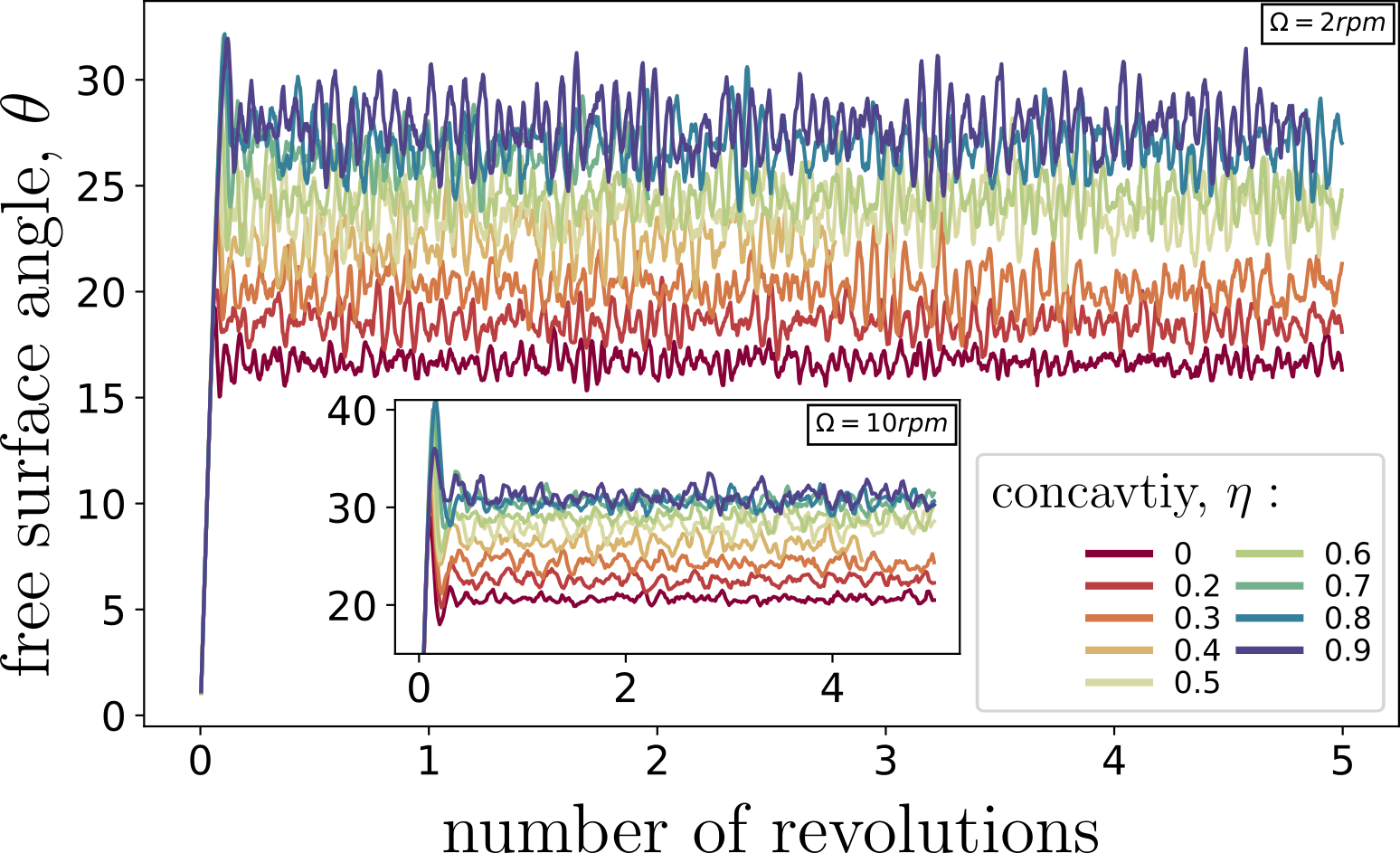}
  \put(-1,61){\bf{(a)}}
  \end{overpic}
  \end{minipage}
  
  \vspace{0.5cm}
  \begin{minipage}[t]{0.9\linewidth}
  \begin{overpic}[width=1\textwidth]{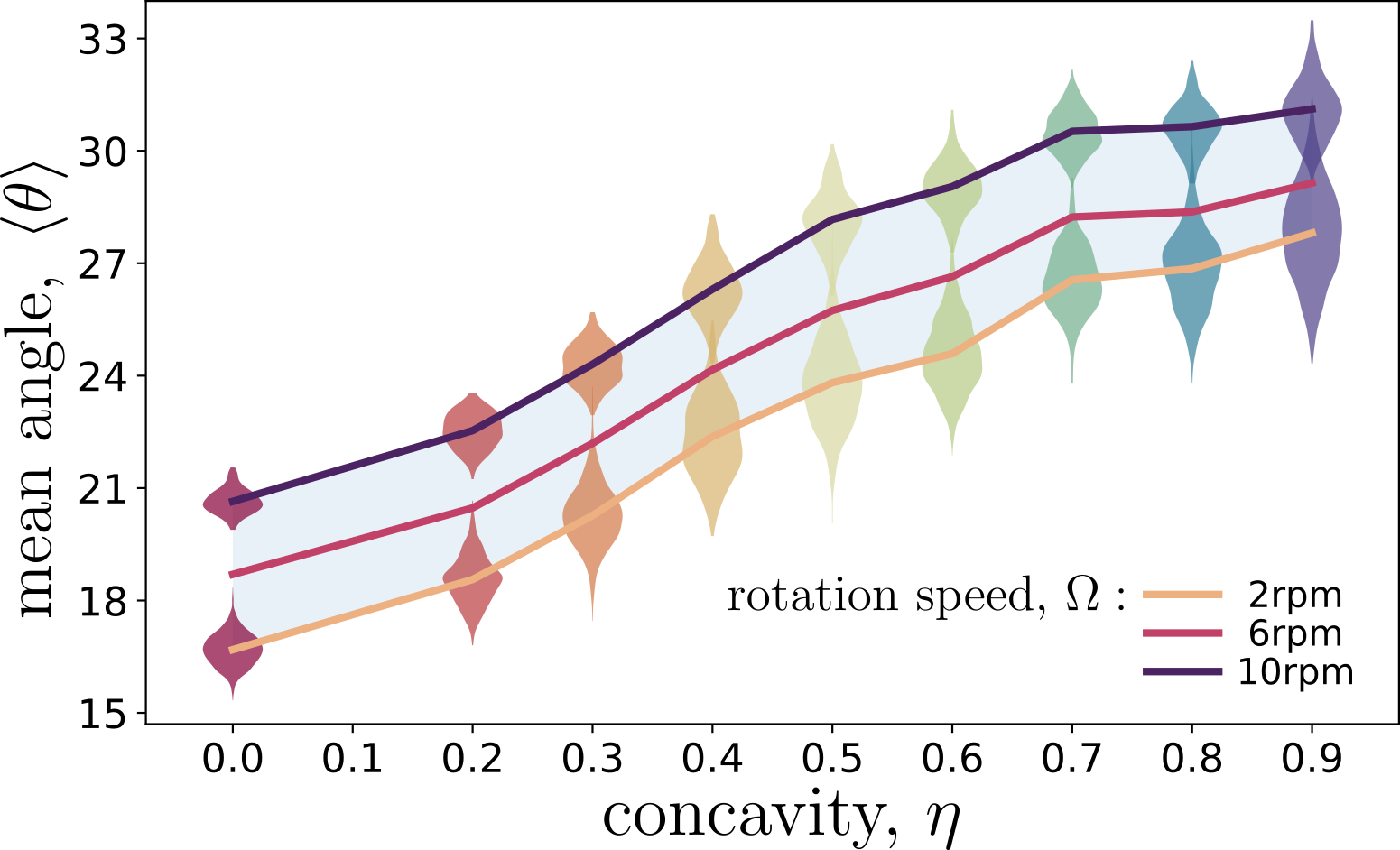}
  \put(-1,61){\bf{(b)}}
  \end{overpic}
  \end{minipage}
\caption{
(a) Evolution of the free surface angle, $\theta$, as a function of the number of revolutions for a drum speed $\Omega=2$~rpm ($\Omega=10$~rpm inset) and different particle shapes.(b) $\langle \theta \rangle$, the value of $\theta$ averaged between revolution $1$ and $5$, as a function of $\eta$ after one revolution at drum speeds $\Omega=2,6,10$~rpm. The histogram of $\theta$ values is also shown vertically for drum speeds $\Omega=2$ and $10$~rpm.}
\label{Macro_mean_num_theta}
\end{figure}

Then, a constant angular velocity $\Omega$ is applied to the drum.
$\Omega$ is varied in $[2,4,6,8,10]$~rpm for each values of $\eta$. 
For every configuration, we make sure that a permanent steady state regime is reached. Practically, when $\Omega$ is applied, all systems slightly dilate. The free surface is then no longer parallel to the reference frame but inclined with an angle $\theta$ as illustrated in Fig. \ref{fig:grans_2d}(b,c). 
Figure \ref{Macro_mean_num_theta}(a) shows the evolution of $\theta$ for $\Omega=2$~rpm ($\Omega=10$~rpm in the inset) as a function of the number of revolutions and for different values of $\eta$. 
As it can be seen, for each pair $(\eta,\Omega)$, on average, $\theta$ remains constant. This evidences that the flow remains in a steady state regime, also called the ``rolling'' flow regime. For every simulation, this steady state is maintained for approximately $5$ revolutions before stopping the simulations.
Moreover, Fig.\ref{Macro_mean_num_theta}(b) shows that $\langle \theta \rangle$ (\textit{i.e.}, the value of $\theta$ averaged between revolution $1$ and $5$) is an increasing, but non-linear, function of $\eta$ that seems to saturate for $\eta>0.7$, for all $\Omega$.

In the following sub-sections, the data we present are averaged over time in the steady state regime. We recall that these averaged profiles are computed within the $\Sigma$ zone at the drum center whose thickness is fixed to $40r$ (see Fig. \ref{fig:grans_2d}(b)). This zone is then vertically divided into layers of thickness equal to $\delta z = 1.5 r$. This choice is motivated by a preliminary sensitivity study of the profiles not presented here. Averages are taken over 480 states saved between the first and the last revolution in the steady state regime.

\subsection{Stress profile prediction}
In this section we aim to validate the predictions of $\sigma_{zz}(z)$ and $\sigma_{xz}(z)$ given by Eq.\ref{eq:Pressure2} and Eq.\ref{Eq_tau2}, respectively. To do so, we first need to evaluate the evolution of the granular stress tensor and the packing fraction as a function of $z$.

\subsubsection{Definitions}
In granular systems, the stress tensor $\bm{\sigma}$ is the sum of three contributions: $\bm{\sigma}=\bm{\sigma}^c+\bm{\sigma}^k+\bm{\sigma}^r$, where each element of this sum refer to the contact, kinetic, and rotational components of the stress, respectively \cite{Renouf2005}. In the dense flow regime, $\bm{\sigma}^k$ and $\bm{\sigma}^r$ are always found to be negligible compared to $\bm{\sigma}^c$. We note that this point is verified for all our simulations. Therefore, it can be assumed that $\bm{\sigma} =\bm{\sigma}^c$. Firstly, for each particle $p$, we build the internal moment tensor $M^p_{ij}= \sum_{c \in p} F^c_i r^c_j$ \cite{Andreotti2013_bk}, where $F^c_i$ is the $i^{\rm{th}}$ component of the force applied on particle $p$ at contact $c$; and $r^c_j$ is the $j^{\rm{th}}$ component of the position vector of the same contact. The sum runs over all contacts $c$ of a particle $p$. Secondly, a Voronoi tesselation, paving the volume (area in 2D) occupied by the grains (see appendix \ref{Appendice_voronoi}), is used to measure an effective volume $V_v^p$ occupied by a particle $p$. Finally, the components of the granular stress tensor $\bm{\sigma}(z)$ at an altitude $z$ is given by \cite{Andreotti2013_bk}: 
\begin{equation}
   \bm \sigma^c(z)= \frac{1}{V_z} \sum_{p \in [z,z+\delta z]} \bm M^p,
   \label{Eq_sigma_z}
\end{equation}
where the sum run over all the particles, $p$, having their center of mass within $[z,z+\delta z]$; and $V_z$ is the sum of the local volumes $V^p_v$ of the corresponding particles.
Finally, along the same line, the packing fraction profile $\phi(z)$ is built using the Voronoi tesselation as:
\begin{equation}
   \phi(z)= \frac{1}{V_z} \sum_{p\in [z,z+\delta z]} V^p,
   \label{Eq_compacity_z}
\end{equation}
where $V^p$ is the volume (area in 2D) of a particle $p$ whose center of mass belong to the slice $[z,z+\delta z]$.

\subsubsection{Profiles}

\begin{figure}
  \centering
  \begin{minipage}[t]{1\linewidth}
  \includegraphics[width=0.95\textwidth]{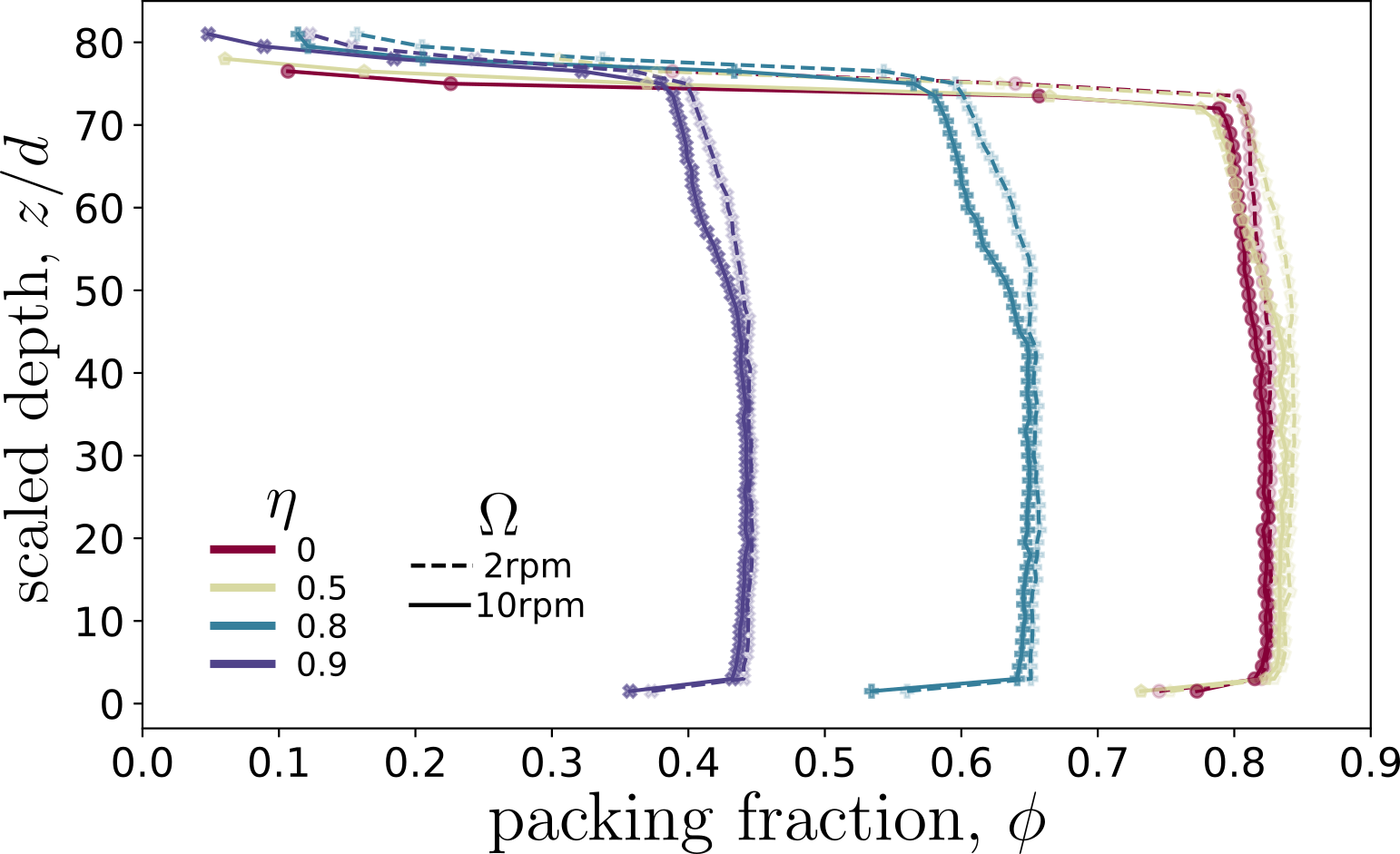}
  \end{minipage}
\caption{Packing fraction profile, $\phi(z)$, for different particle convexity $\eta$, for $\Omega=2$ (dashed line) and $\Omega=10$ (plain line).}
\label{fig:packing_fraction}
\end{figure}

Figure \ref{fig:packing_fraction} shows the packing fraction profiles, $\phi(z)$, for the slowest and fastest angular velocities simulated in this study, for different $\eta$.
As a general observation, the variation of $\phi(z)$ is in good agreement with the description given in Sec. \ref{Sec_Model_velocity_fluidity}. Indeed, $\phi(z)$ is nearly uniform in the bulk and equal to $\phi_s$ until it declines nearly linearly in the flowing layers, from $\phi_s$ at $z_s$ to $\phi_c$ at $z_c$ and diverges close to the free surface, \textit{i.e.}, for $z>z_c$. Yet, close to the drum boundaries, the packing fraction is also found to diverge within a small region one or two grains diameters thick.
These profiles also evidence an expansion of the assemblies at large $\eta$-values -- $\phi$ declines as $\eta$ increases. An expansion is also noticeable with $\Omega$ but only in the flowing zone.

\begin{figure}
  \centering
  \begin{minipage}[t]{0.9\linewidth}
  \begin{overpic}[width=1\textwidth]{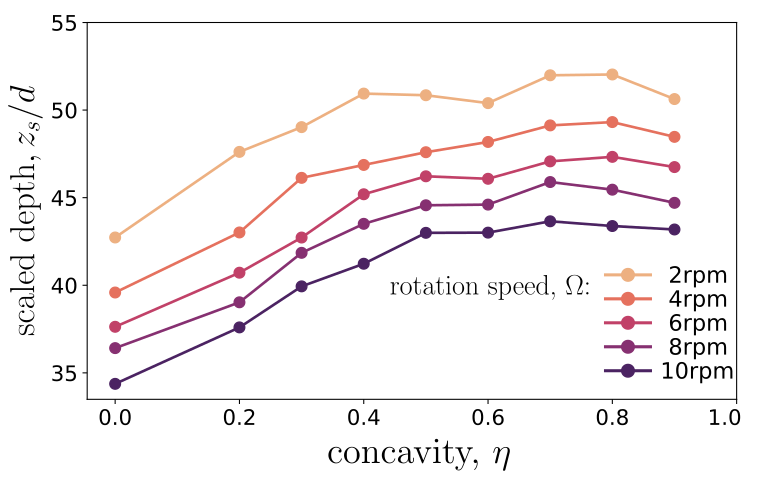}
  \put(-1,61){\bf{(a)}}
  \end{overpic}
  \end{minipage}
 
  \hspace*{0.3cm}
  
  \begin{minipage}[t]{0.9\linewidth}
  \begin{overpic}[width=1\textwidth]{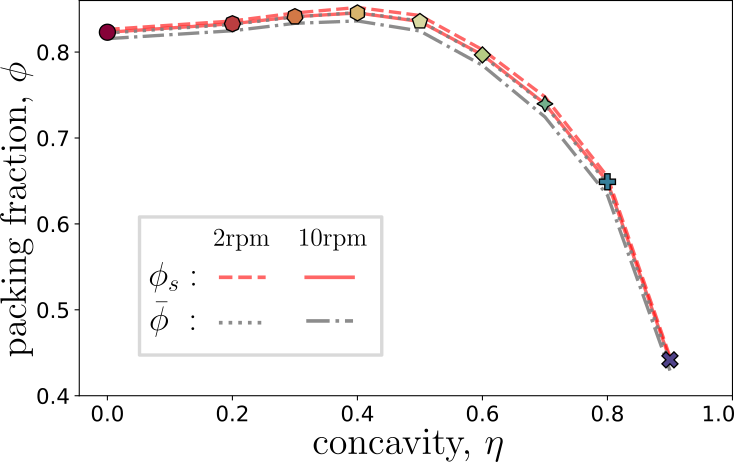}
  \put(-1,61){\bf{(b)}}
  \end{overpic}
  \end{minipage}
\caption{(a) Evolution of $z_s$ (the transition depth between static and quasi-static zones) as a function of the shape parameter $\eta$. (b) Evolution of $\phi_s$ (the packing fraction averaged in the solid-like zone) and $\bar \phi$ (the packing fraction averaged in the liquid zone, between $z_s$ and $z_c$) as functions of the shape parameter $\eta$ for $\Omega=2$ (dashed line) and $\Omega=10$ (plain line).}
\label{fig:meanpacking_fraction_z}
\end{figure}

Figure \ref{fig:meanpacking_fraction_z}(a) shows the variation of $z_s$ as a function of $\eta$ for the different rotation speed $\Omega$. This quantity is measured from a linear fit of the packing fraction profiles as explained in Sec.\ref{Sec_Model_velocity_fluidity} and illustrated in Fig.\ref{fig_3}. We recall that it is a characteristic depth of the system that feeds the equation of velocity profile (Eq.\ref{eq:shear_rate}). As a general observation, $z_s$ evolves non-linearly with $\eta$: it first increases and, beyond $\eta=0.4$, remains relatively constant. Meanwhile, $z_s$ declines as $\Omega$ increases but conserves its trend as a function of $\eta$.
Figure \ref{fig:meanpacking_fraction_z}(b) shows the variation of both quantities $\phi_s$ and $\bar \phi$ as functions of $\eta$ for $\Omega=2$ and $10$~rpm. We recall that $\phi_s$ is the packing fraction averaged in the solid-like zone, while $\bar \phi$ is the packing fraction averaged in the liquid zone between $z_s$ and $z_c$. Both quantities fairly coincide regardless of the rotation speed, $\Omega$.
Interestingly, these packing fractions slightly increase with $\eta$, reach a maximum for $\eta=0.5$, and then quickly decline as $\eta$ further increases. It is worth noting that a similar non-monotonous behavior has been observed in previous works with elongated \cite{CEGEO2012,Daniel2017}, angular \cite{CEGEO2012,Emilien2013,Athanassiadis2014} or slightly non-convex \cite{Emilien2012,Emilien2013,Athanassiadis2014} grains in both two and three dimensions. This enlights a generic feature also valid for highly non-convex grains when the shape deviates continuously from a circular shape. 

\begin{figure}
  \begin{minipage}[t]{0.9\linewidth}
  \begin{overpic}[width=1\textwidth]{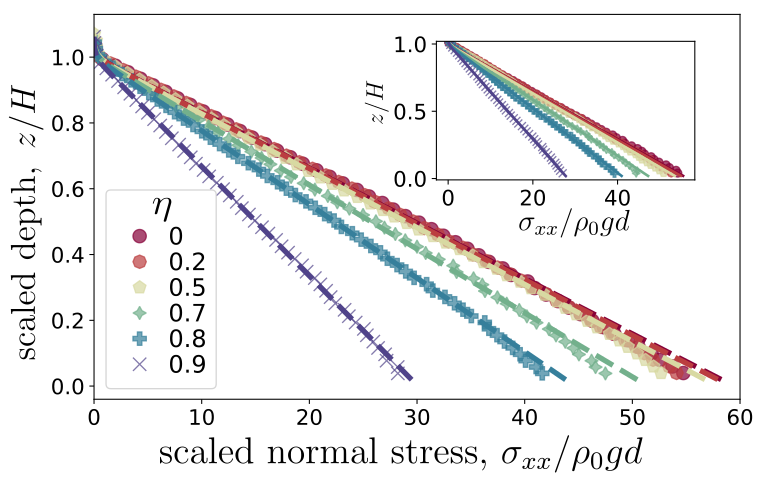}
  \put(-1,61){\bf{(a)}}
  \end{overpic}
  \end{minipage}
  
  \hspace*{0.3cm}
  
  \begin{minipage}[t]{0.9\linewidth}
  \begin{overpic}[width=1\textwidth]{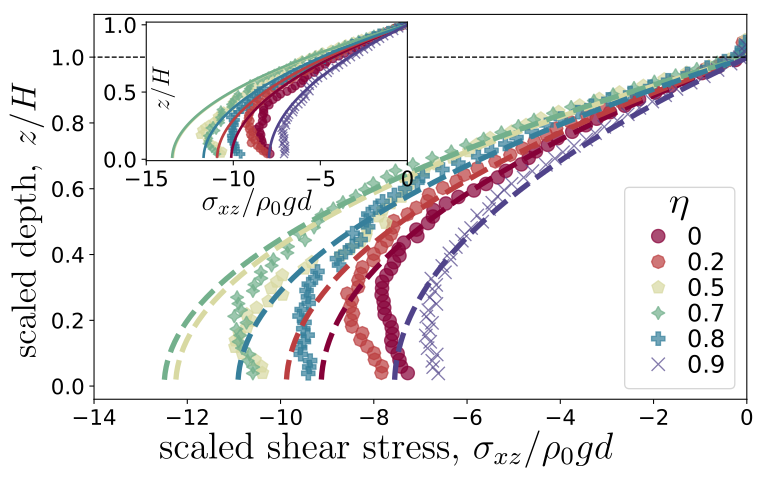}
  \put(-1,61){\bf{(b)}}
  \end{overpic}
  \end{minipage}
\caption{Normal stress profile $\sigma_{zz}$ (a) and The shear stress profile $\sigma_{xz}$ (b) for different values of $\eta$ and for $\Omega=2$~rpm and $\Omega=10$~rpm (inset). The simulation data are presented as symbols and the model predictions (Eq.\ref{eq:Pressure2} and Eq.\ref{Eq_tau2}, respectively) by dashed lines.}
\label{fig:simus_stress}
\end{figure}

Then, Fig. \ref{fig:simus_stress} displays the normal stress $\sigma_{zz}(z)$ (a) and shear stress $\sigma_{xz}(z)$ (b) profiles for different values of $\eta$, and for different loading speed $\Omega=2$ and $10$~rpm. We observe that $\sigma_{zz}(z)$ decreases linearly with $z$ and decreases with $\eta$ but more complexly. On the contrary, from the top layer, $\sigma_{xz}(z)$ first decreases with depth, and then saturates close to the drum border on a distance comprised between $0.2H$ and $0.3H$ depending on $\eta$. 
When increasing this latter parameter, the shear stress first increases (in absolute value), but beyond $\eta=0.5$ it goes back to values slightly lower than that of disk assemblies, for $\eta=0.9$. This non-monotonous variation of the shear stress with $\eta$ recalls that of the packing fraction observed just above and will be discussed below.

The stress profiles $\sigma_{zz}(z)$ are outstandingly well approximated by Eq.\ref{eq:Pressure2} for all pairs $(\eta,\Omega)$ used in this study. The prediction of $\sigma_{xz}(z)$  given by Eq.\ref{Eq_tau2} also nicely reproduces the main variations of the shear stress profiles across the drum for all pairs $(\eta,\Omega)$, from the free surface up to deep in the solid-like phase. It correctly captures the constant shear stress profile observed close to the border. The minor mismatch with the numerical data mainly comes from the fact we impose $\sigma_{xx}=\sigma_{zz}$ to write Eq.\ref{Eq_tau2}. This is well verified in bulk, but not necessarily close to the border, consistently with previous works \cite{Renouf2005, Povall2018}. 
For conciseness and clarity, a punctual discussion is added in appendix \ref{A_Correction_stress_with_xx_yy} on this specific aspect.

\begin{figure}
  \begin{minipage}[t]{0.95\linewidth}
  \begin{overpic}[width=1\textwidth]{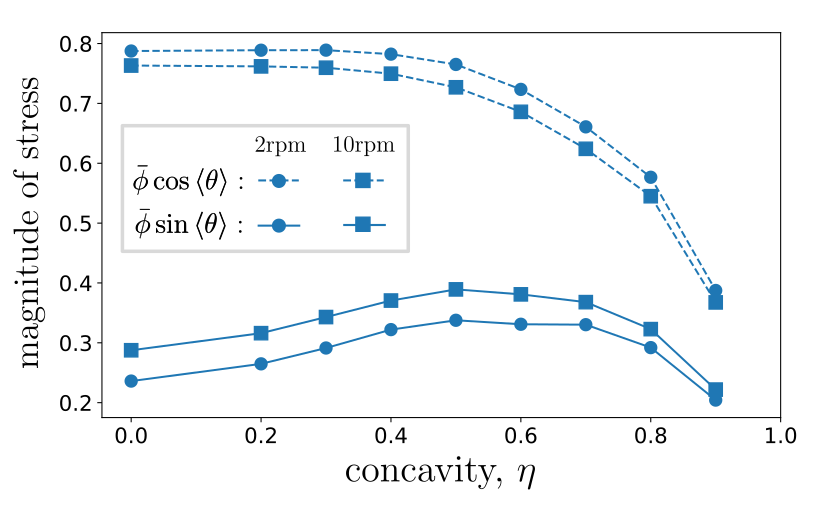} \end{overpic}
  \end{minipage}
\caption{Evolution of the slopes $\bar \phi \cos \langle \theta \rangle$ (dashed line) and  $\bar \phi \sin \langle \theta \rangle$ (full line) of the corresponding normal (Eq. \ref{eq:Pressure2}) and shear (Eq.\ref{Eq_tau2}) stress profiles, as functions of $\eta$, for $\Omega=2$~rpm (circle) and $\Omega=10$~rpm (squares)}
\label{fig:pre_factor_stress}
\end{figure}

Finally, following of Eqs. \ref{Eq_tau2} the non-monotonous variation of $\sigma_{xz}$ as a function $\eta$ can be better understood from the combined contributions of the evolution of $\bar \phi$ and $\langle \theta \rangle$ with this same parameter, $\eta$. Indeed, for small $\eta$-values both $\sin \langle \theta \rangle$ and $\bar \phi$ increase, which explains the increase of $\sigma_{xz}$. On the contrary, for $\eta>0.5$ the rapid decrease of $\bar \phi$ with $\eta$ induces a decrease of $\sigma_{xz}$. As a result, the product $\bar \phi \sin \langle \theta \rangle$ increases first with $\eta$ but declines from $\eta=0.5$ up to values lower than that of disk packings (see Fig.\ref{fig:pre_factor_stress}).
The same holds for the variation of $\sigma_{zz}$ with $\eta$. In this latter case, the small increases of $\bar \phi$ at small $\eta$ is compensated by the decrease of $\cos \langle \theta \rangle$. On the contrary, at larger $\eta$, both $\bar \phi$ and $\cos \langle \theta \rangle$ decline, so that their product continuously declines with $\eta$ (see Fig.\ref{fig:pre_factor_stress}).

\subsection{Velocity profile prediction and fluidity}
In this section, we test the predictions of velocity profiles given by Eq.\ref{eq:shear_rate}. We discuss the effect that both, the grain shape, $\eta$, and the rotation speed, $\Omega$, have on the thickness, $\lambda$, of the inertial flow zone. 

\begin{figure}
  \begin{minipage}[t]{1\linewidth}
  \includegraphics[width=0.95\textwidth]{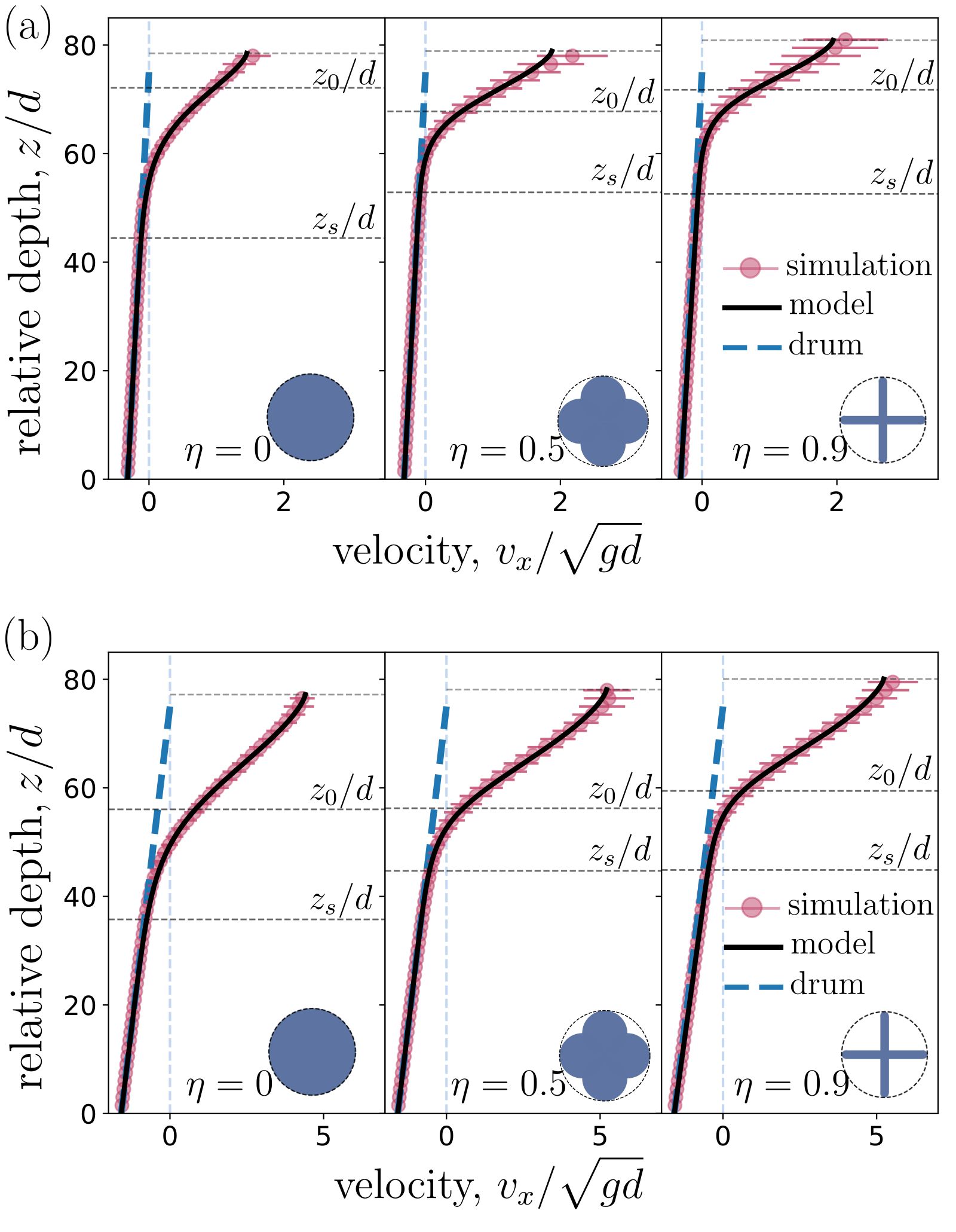}
  \end{minipage}
\caption{Velocity profiles measured in the numerical simulations along the $e_x$ direction at drum speed $\Omega=2$~rpm (a) and $\Omega=10$~rpm (b) for different $\eta$. In both cases (a) and (b), the plain line represents the velocity profile of the model (Eq.\ref{eq:shear_rate}), and, the blue dashed line represents the velocity profiles of the drum. The simulation data are shown as scatter plots.
}
\label{fig:simus_velocity}
\end{figure}

Fig.\ref{fig:simus_velocity} displays the $x$-velocity profiles, $v_x$, averaged in the slice $\Sigma$, for $\eta \in[0,0.5,0.9]$ and at drum speeds $\Omega=2$~rpm (a) and $\Omega=10$~rpm (b). Symbols are direct measurements of the numerical simulations while the plain lines are the predictions given by Eq.\ref{eq:shear_rate}. We observe that for every shape and drum speed, a solid-like behavior is displayed deep in the packing. This corresponds with the speed profile following the straight blue dashed lines in Fig. \ref{fig:simus_velocity}. Then, in the upper layers, the velocity profiles increase and split from the drum velocity when both, $\eta$ and $\Omega$ increase.

\begin{figure}
  \centering
  \begin{minipage}[t]{0.95\linewidth}
  \includegraphics[width=0.95\textwidth]{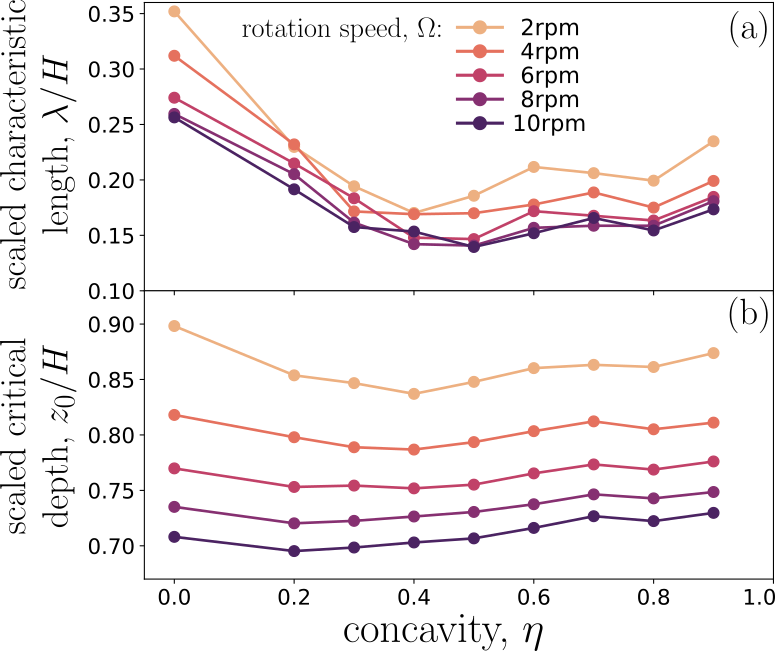}
  \end{minipage}
  \caption{Variations of the scaled characteristic thickness of the inertial zone, $\lambda/H$ (a), and of the scaled critical depth, $z_0/H$ (b)  plotted as functions of the concavity, $\eta$, for different drum speeds, $\Omega$.
    }
  \label{fig:paras_simus}
\end{figure}

The prediction of the relative velocity profile, given by Eq. \ref{eq:shear_rate}, is shown in plain lines in Fig. \ref{fig:simus_velocity} by fitting the free parameters $(k,\xi,\lambda)$. We remember that $z_s$ is fixed from the packing fraction profiles curves (see Fig.\ref{fig:packing_fraction}). We see that the theoretical prediction is outstanding for every $\eta$ and $\Omega$. 
The way the fitting is carried out is first by fitting the ``universal'' constant $k$ by minimizing the sum of residuals for all the fits, corresponding with all the simulations. We get $k=10$. Then, the two unknowns $\xi$ and $\lambda$ are solved using the two equations Eq.\ref{eq:shear_rate} (for the velocity profile) and Eq.\ref{eq:Fphi} (for the fluidity profile) simultaneously. Doing so, we find that $\xi=2$, independently of $\eta$ and $\Omega$ (see also Fig.\ref{Fig_fluidity_fF} in Appendix \ref{Appendice_f_F}). On the contrary, as shown in Fig.\ref{fig:paras_simus}(a), $\lambda$ varies significantly with $\eta$ and in a lower proportion with $\Omega$. More precisely, $\lambda$ first decreases with $\eta$ and then plateau from $\eta=0.4$ while it continuously increases with $\Omega$. From $\lambda$ we can then compute the transition depth between the quasistatic and inertial flow, $z_0$, Interestingly, as shown in Fig.\ref{fig:paras_simus}(b), this latter remains independent of $\eta$ but decreases with $\Omega$.

The strength of the model we propose in this paper lies not only in the correct modeling of the velocity, density, and stress profiles for any grain shape and loading speed but also in the ability to predict the thickness of the different flow layers and their depth.

\section{\label{sec:EXP}EXPERIMENTAL VALIDATION}

In this section, we develop a series of grain flow experiments in a rotating drum. We used molded rigid cross-shaped 3D grains with different levels of convexity to test our theoretical velocity model experimentally.

\subsection{Experimental set-up}
\begin{figure*}
  \centering
  \includegraphics[width=0.8\textwidth]{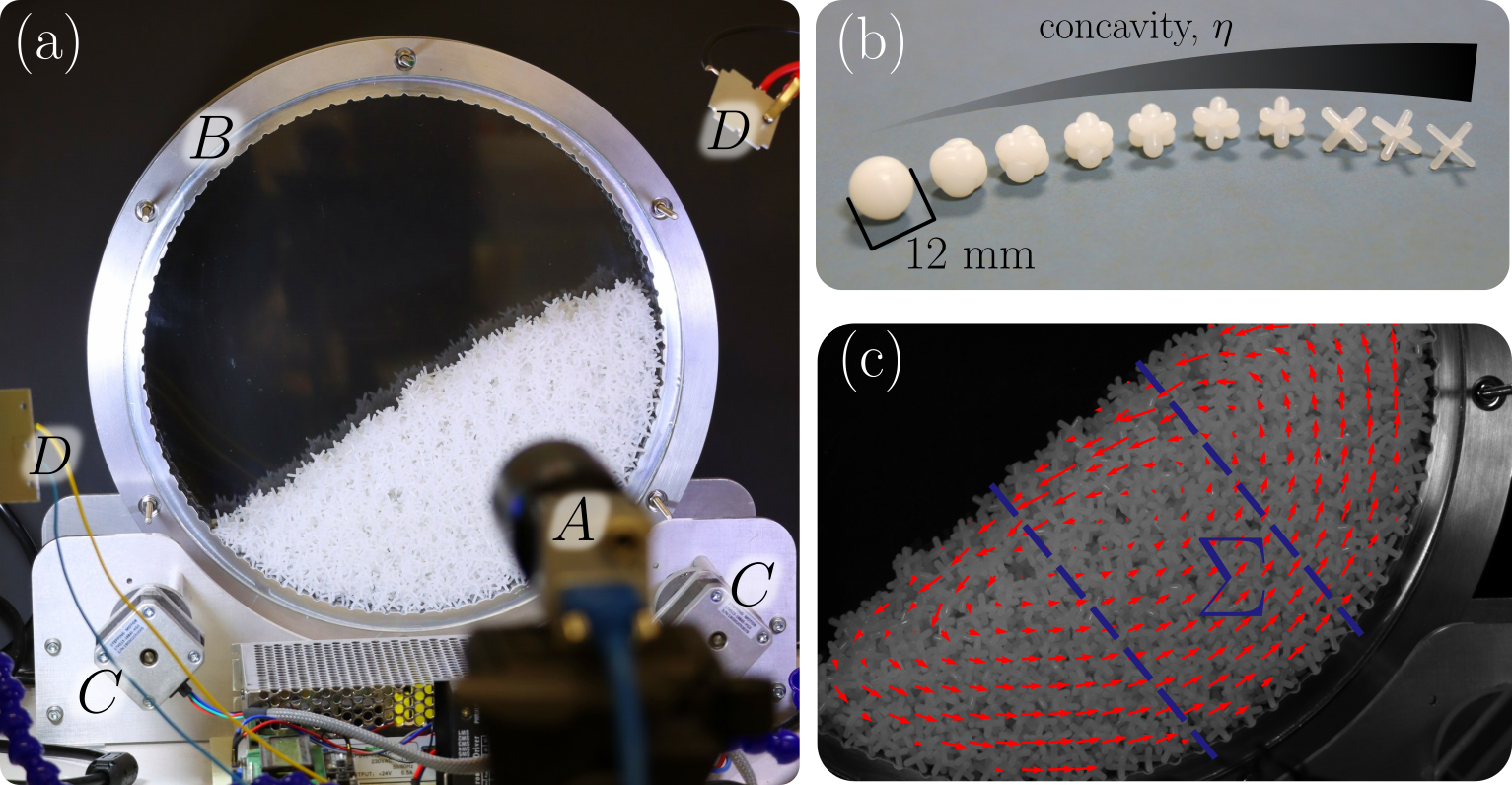}
\caption{(a) Picture of the experimental setup: $A$ camera, $B$ drum, $C$ stepper motors, $D$ lighting. (b) Particles with varying shapes, ranging from spherical ($\eta = 0$\%) to highly concave ($\eta = 87.5$\%). When the concavity, $\eta$, increases, the spherocylinder diameter gradually decreases from $12$~mm to $1.5$~mm. (c) Velocity field obtained from Particle Image Velocimetry (PIV) on top of flow picture for particles with concavity $\eta = 79$\%. The dark blue dotted line represents the study slice $\Sigma$, which has a width of $6$ particle diameters ($7.2$~cm).
}
\label{fig_exp_1}
\end{figure*}

We use The rotating drum shown in  Fig.\ref{fig_exp_1}(a). It is a homemade device with an inside diameter of $28$~cm and a depth of $5$~cm. It is filled with monodisperse particles of shape varied from spherical to highly concave as presented in fig.\ref{fig_exp_1}(b). All the particles are circumscribed in a sphere of diameter $d=12$~mm. The concave ones consist of $3$ spherocylinders that extend toward the direction of the faces of a regular cube. The radius of the spherocylinders, $r_0$, gradually increases from $0.75$~mm to $6$~mm (sphere). These particles are made by injection molding of high-density polyethylene (HDPE). A dedicated mold producing clusters of $20$ particles has been custom made \cite{particle_making}. This material's friction coefficient is very low, close to $0.1$, and its Young modulus is quite high, around $1$~GPa. This makes the particles rigid and slippery.
The concavity parameter $\eta$ (see Eq.\ref{Eq_concavity_parameter}) is 
varied such $\eta \in[0,0.33, 0.5, 0.58, 0.67, 0.71, 0.75, 0.79, 0.83, 0.875]$.

The drum has smooth transparent glass on both axial sides, and the inner radial side is regularly raised to prevent sliding. A constant volume of $900$~mL of particles is loaded into the drum for each experiment. The drum is illuminated with LED light, and a camera \cite{ref_camera} is positioned perpendicular to the front glass to image the system at a high frequency ($60$ frames per second). The rotation speed of the drum is controlled by tuning the speed of two synchronized stepper motors below it (see fig.\ref{fig_exp_1}(a)). We tested three angular velocities for the drum speed: $\Omega \in[1.93, 2.91, 4.83]$~rpm. For each couple $(\eta, \Omega)$ the drum is first rolled during $2$ min to be sure to reach the steady state. Then, image recording is launched during $5$~min. 

The mean behavior for each set of parameters $\eta$ and $\Omega$ is obtained by averaging over three independent data sets. This means that each experiment is repeated $3$ times, so in total $33$ experiments are performed. Image analysis shows that the flows stay in the rolling regime for these pairs of parameters:
$(\eta,\Omega)\in \{ ([0,..,0.58],1.93) \; ; \; ([0,..,0.67],2.91) \; ; \; ([0,..,0.79],4.83) \}$. In the following, only these data sets are used.

\begin{figure}
  \centering
  \includegraphics[width=0.95\linewidth]{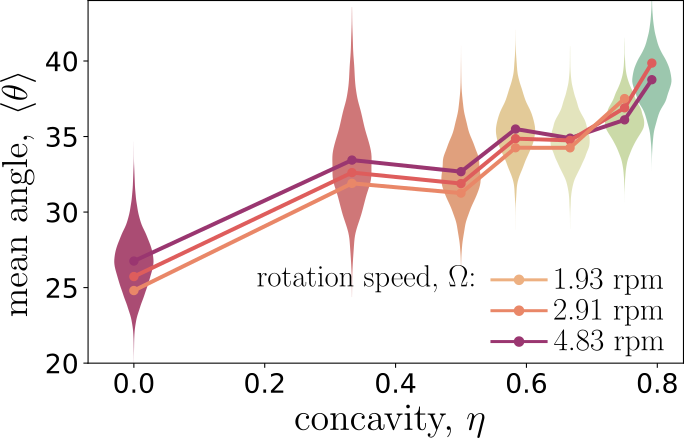}
\caption{Variation of the time-average free surface angle $\langle \theta \rangle$ as a function of $\eta$ and for different drum speed $\Omega \in[1.93, 2.91, 4.83]$~rpm.
The vertical histogram represents the distribution of the surface angles for a rotational speed $\Omega=4.83$~rpm.}
\label{fig_exp_angle}
\end{figure}

Figure \ref{fig_exp_angle} shows $\langle \theta \rangle$ as a function of $\eta$ for the three rotation speed $\Omega$. Consistently with the numerical simulations (see Fig.\ref{Macro_mean_num_theta}(b) for recall), $\langle \theta \rangle$ is an increasing function of both $\eta$ and $\Omega$.

\subsection{Velocity profile}
\begin{figure}
  \centering
  \includegraphics[width=0.95\linewidth]{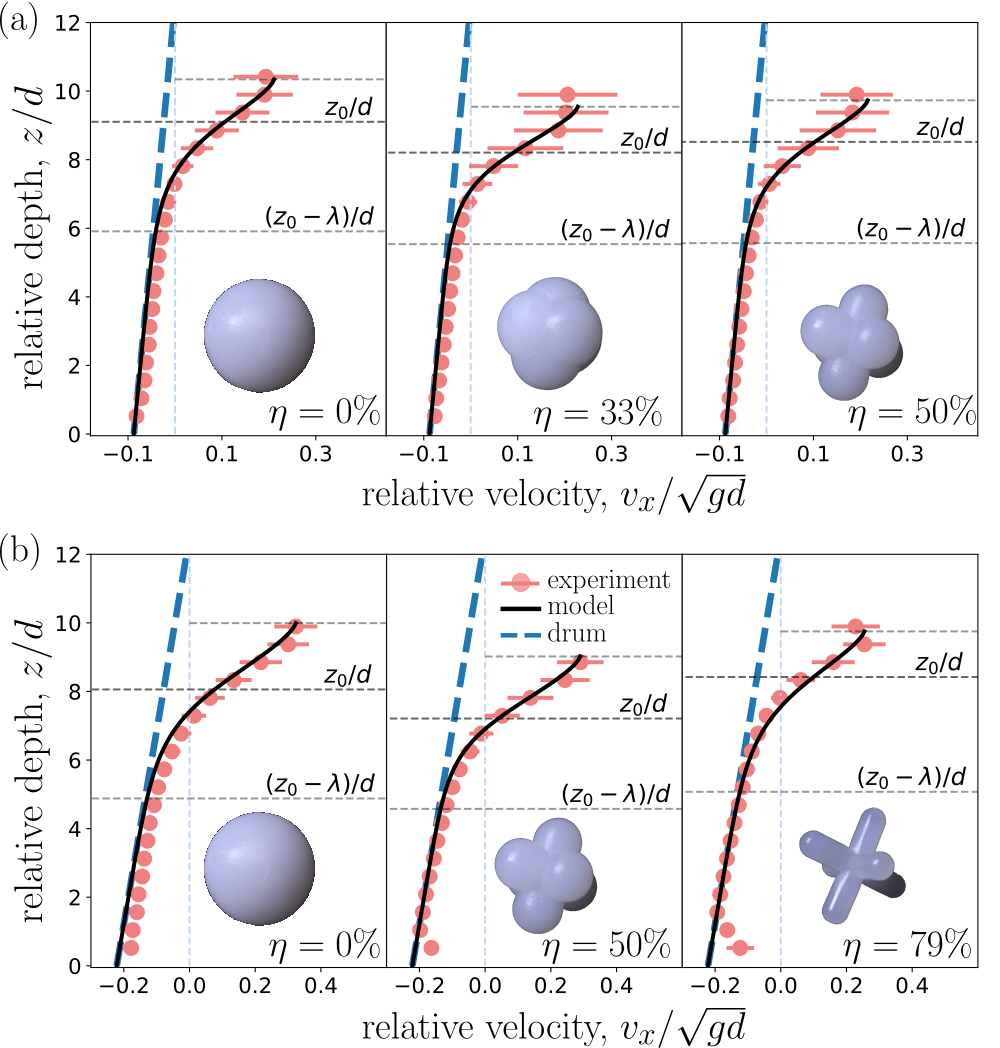}
\caption{Velocity profiles measured experimentally along the $e_x$ direction at drum speed $\Omega=1.93$~rpm (a) and $\Omega=4.83$~rpm (b) for different $\eta$. In both cases (a) and (b), the plain line represents the velocity profile of the model (Eq.\ref{eq:shear_rate}), and, the blue dashed line represents the velocity profiles of the drum. The experimental data are shown as scatter plots.}
\label{fig_exp_2}
\end{figure}

From the images, we deduce the instantaneous velocity field, $\pmb{v}$, using Particle Image Velocimetry (PIV) as shown in fig.\ref{fig_exp_1}(c). Then, we decompose this velocity field into directions parallel ($\pmb{e}_x$) and perpendicular ($\pmb{e}_z$) to the free surface. The so obtained $x$-velocity component, $v_x$, is averaged in time within the slice $\Sigma$ to obtain the velocity profiles $v_x(z)$ (see Fig.\ref{fig_exp_2}). We recall that the blue dashed line shows the linear evolution of the drum velocity as a function of the depth. The plain black line is the velocity profile fitted by the analytical model (see Eq. \ref{eq:shear_rate}). 

As we can see, it is clear that the velocity profiles obtained experimentally follow the same trends as those obtained numerically. It should be noted that, from experiments, we cannot measure the evolution of the packing fraction. So we cannot determine the parameter $z_s$ {\it a priori}, and Eq.\ref{eq:Fphi} cannot be used based on the evolution of the solid fraction. Thus, contrary to the numerical case, Eq.\ref{eq:shear_rate} is therefore based on $4$ parameters that have to be determined. Nevertheless, based on the numerical simulations we can fix $\xi=2$ since we have seen that this parameter, a threshold value to ensure that $f(z)=F(\phi(z))$, is independent of both, $\Omega$ and $\alpha$. Then, $k$ is fixed to $1$ using the same strategy as in the numerical case, \textit{i.e.}, by first minimizing the sum of all residuals for all fits. Once $k$ and $\xi$ are determined, the two last parameters are fitted for each experiment.

\begin{figure}
  \includegraphics[width=0.8\linewidth]{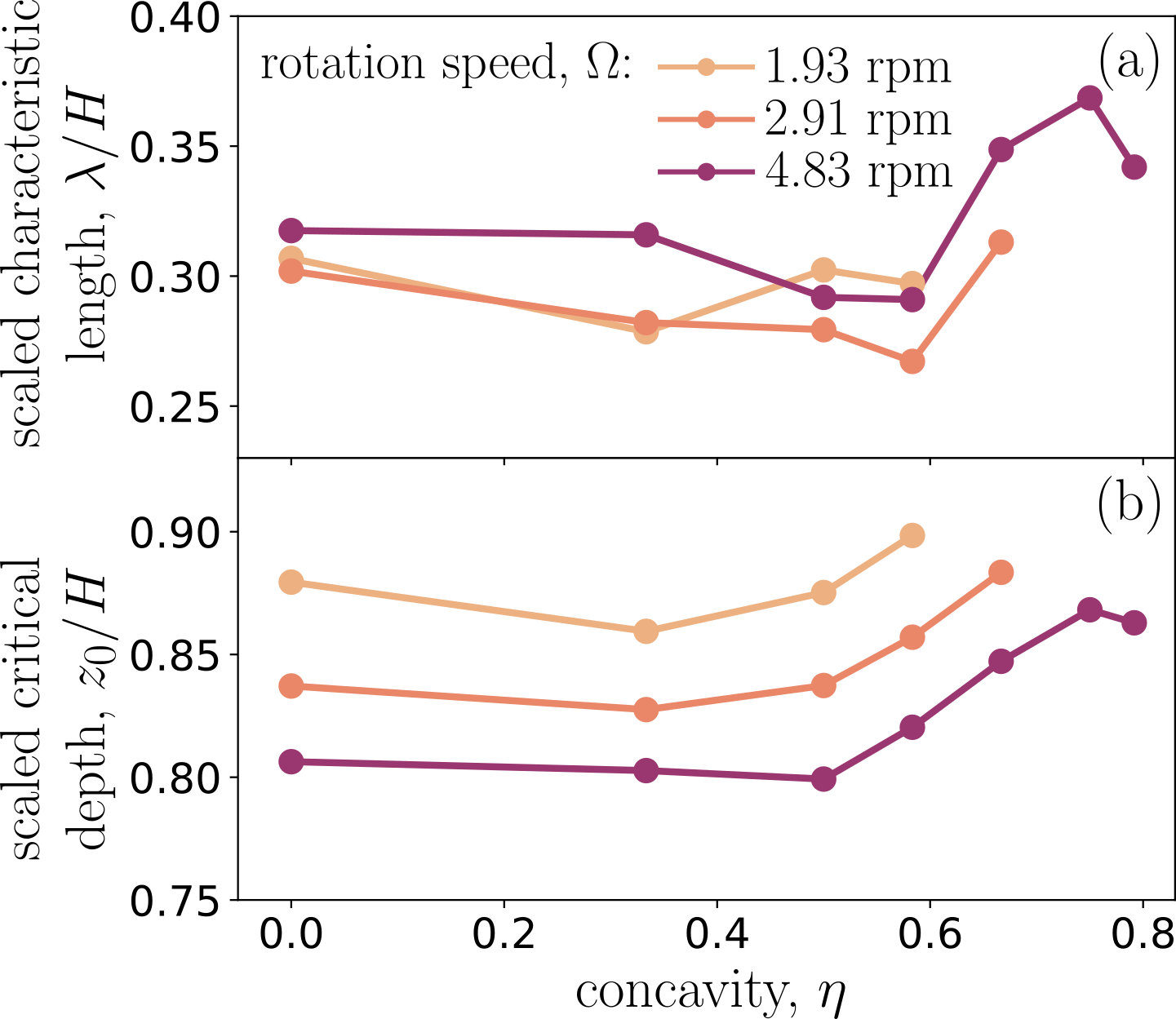}
  \caption{ Variations of the scaled characteristic thickness of the inertial zone, $\lambda/H$ (a), and the scaled critical depth, $z_0/H$ (b), plotted as functions of the concavity, $\eta$, for different drum speeds, $\Omega$ in the experiments.  }
  \label{fig_exp_3}
\end{figure}

Figure \ref{fig_exp_3} displays the variations of the scaled characteristic thickness of the inertial zone, $\lambda/H$ (a), and of the scaled critical depth, $z_0/H$ (b). Both are plotted as functions of the concavity, $\eta$, for different drum speeds, $\Omega$. Consistently with the numerical results, the critical depth, $z_0$, decreases when the rotating speed, $\Omega$, increases. Yet, for a given $\Omega$, it is almost constant as long as $\eta$ is low enough, and increases (faster as in the 2D numerical simulations) with this latter parameter for values above $0.5$. 
The thickness of the inertial zone flow, $\lambda/H$, is quite independent of the rotation speed $\Omega$, as already observed numerically. On the contrary, its variation as a function of $\eta$ in the experiments slightly differs from the simulations. 
Specifically, $\lambda/H$ decreases less steeply than in the numerical simulations. We also observe an increase for the highest $\eta$ values, whereas it tends to plateau in the simulations. Finally, in addition to the very similar trends observed in the variation of these parameters as a function of $\eta$ and $\Omega$, we note that similar (or at least very close) values are obtained in particular for the $\eta=0$ case, which is the most ``comparable'' between the numerical and experimental data. In this latter case for example, for $\Omega=2$~rpm, $\lambda/H$ is close to $0.3$ in the 3D experiments and close to $0.35$ in the 2D simulations. The same holds for $z_0/H$ close to $0.90$ in both approaches.

\section{\label{sec:CONCLUSION}CONCLUSION}

In this study, we have presented an analytical model to describe the evolution of the steady granular flow in a rotating drum. Our approach relates the shear stress $\tau$ to the angle of repose of the free surface $\theta$ and introduces the shear rate $\dot{\gamma}$ and packing fraction $\phi$ into the generalized Bagnold equation through the ``fluidity'', $f$. We have found $f$ to depend only on the packing fraction $\phi$. 

To test our model, we designed a series of 2D numerical simulations (based on a discrete element method) and 3D experiments involving grain flow in a rotating drum. In addition, to test the robustness of the theoretical model, we went as far as considering the case of granular systems made of very non-convex grains. Grains with a very non-convex shape have the peculiarity of being entangled and thus present flow patterns that can be radically different from those of convex grains. Thus they constitute a fascinating extreme validation case.

Our numerical and experimental results confirm the validity of the model for steady granular flows and its ability to accurately describe the velocity profile, regardless of the grain shapes. Furthermore, our simulations confirm the model's efficacy in predicting stress and packing fraction evolutions within the assembly. The combination of experiments and simulations compared to the model demonstrates the power of an approach introducing the concept of ``fluidity'' in the analytical study. Additionally, we provide an explicit form for this fluidity function ($F(\phi(z))$) for cross-validation. 

In our analytical model, the ``fluidity'' can be viewed as an approximate solution to the diffusion equation based on Landau's theory. To avoid the artificial building of a source term in the diffusion equation, we introduce a  constant $k$ into the ``fluidity''. This ``fluidity'' is then incorporated into the Bagnold scaling with the Prandtl mixing length scale to correct the mixing length. The successful combination of Landau theory and Prandtl mixing length in a macroscopic system primarily governed by inertia provides us with a more intuitive understanding of phase transitions and fluid mechanisms. This hyperbolic logistic function appears in many statistical problems related to phase transitions and provides a highly approximate solution to bridge the different states of the order parameter. It is widely used not only in physical phase transitions but also in ecological \cite{pearl1920rate}, epidemiologic \cite{lee2020}, and chemical \cite{Xi2018} phase transitions. We apply a hyperbolic tangent function to the order parameter ``fluidity'' in granular flow, and the results of our fit exceed expectations. This approach may have potential applications in similar rheological systems of amorphous glassy materials, such as concentrated emulsions, pastes, and molecular glasses.

In addition, our analytical framework incorporates most of the key physical quantities in particle flow. In particular, we evidenced a critical height $z_s$, which marks the transition between the solid and fluid regimes, and a critical thickness $\lambda$ characterizing the inertial flow zone. But even more remarkably, our approach, through the concept of fluidity, highlights a characteristic length scale specific to the grain shape, $l$, which is explicitly linked to the critical thickness by the relationship we establish $\lambda = \xi \sqrt{8} l $. In other words, our model explicitly incorporates the grain shape through this intrinsic grain length scale, $l$.

In this article, we have focused on the modeling of the macroscopic flow. However, a lot of work remains to be done to characterize the microscopic properties, particularly in the presence of very non-convex grains. As mentioned above, these grains can become entangled and most likely form complex clusters connected by multiple contacts. Such a description, with a view to future modeling, remains an open research topic and will be the subject of a detailed study in forthcoming publications.

\bigskip

The authors acknowledge financial support from ANR MICROGRAM (ANR-20-CE92-0009). 
We also acknowledge the support of the High-Performance Computing Platform MESO@LR.
Finally, we would like to thank Gille Camp, Gilles Genevois, and Quentin Chapelier for their technical help in setting up the experimental device and Benjamin Gallard and Sylvain Buonomo for their help in making particles.

\newpage

\appendix
\section{Macroscopic friction profile}
\label{A_profil_mu}
\begin{figure}[H]
  \begin{minipage}[t]{0.9\linewidth}
  \begin{overpic}[width=1\textwidth]{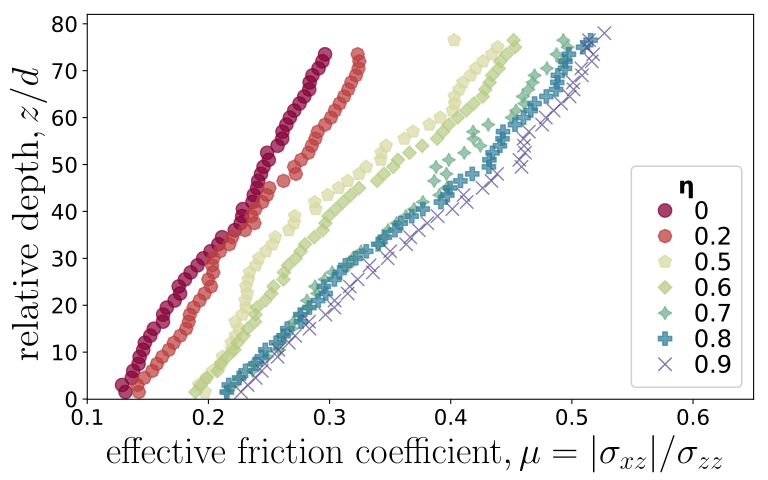}
  \put(-1,61){\bf{(a)}}
  \end{overpic}
  \end{minipage}
  
  \vspace{0.5cm}
  \begin{minipage}[t]{0.9\linewidth}
  \begin{overpic}[width=1\textwidth]{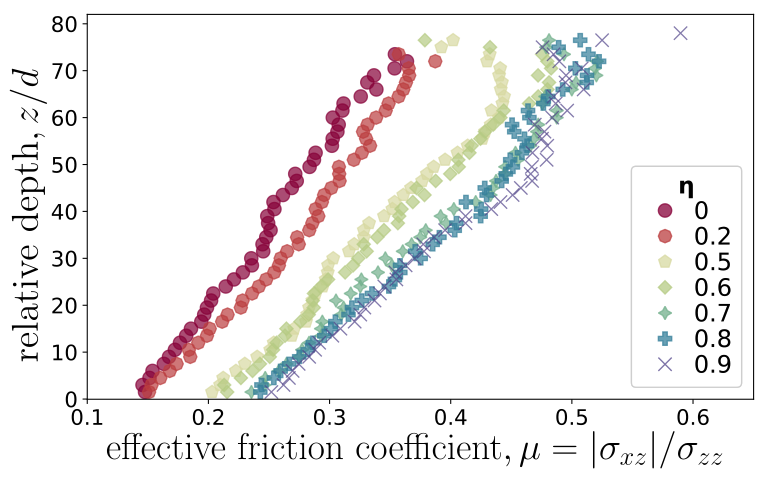}
  \put(-1,61){\bf{(b)}}
  \end{overpic}
  \end{minipage}
\caption{
Effective friction coefficient, $\mu$. (a) $\mu$ is plotted for different values of $\eta$ and for $\Omega=2$~rpm. (b) $\mu$ is plotted for different values of $\eta$ and for $\Omega=10$~rpm}
\label{Effective_friction_coefficient}
\end{figure}

Figure \ref{Effective_friction_coefficient} shows the variation of the macroscopic friction profile $\mu(z)$ as a function of $z$. As we can see, $\mu(z)$ evolves approximately linearly with $z$ in our numerical simulations (as also reported in \cite{Lin2020}). 

\section{Numerical approximation for solving Eq.\ref{eq:Fphi}}
\label{Appendice_f_F}
\begin{figure}[H]
  \begin{minipage}[t]{0.9\linewidth}
  \begin{overpic}[width=1\textwidth]{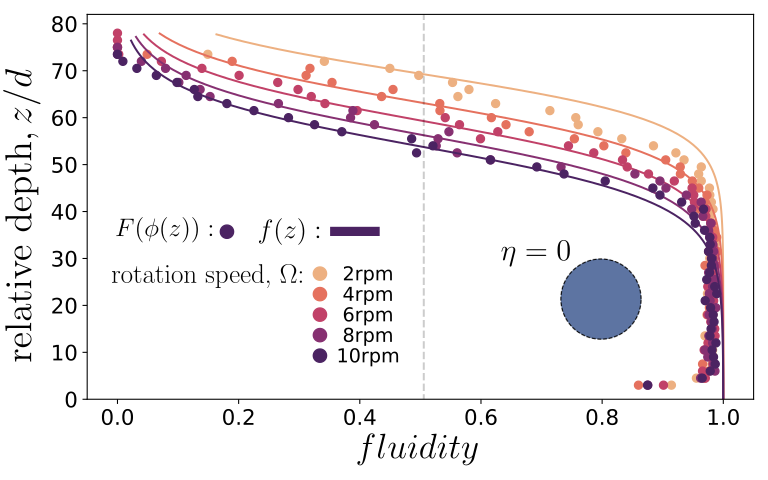}
  \put(-1,61){\bf{(a)}}
  \end{overpic}
  \end{minipage}
  
  \vspace{0.5cm}
  \begin{minipage}[t]{0.9\linewidth}
  \begin{overpic}[width=1\textwidth]{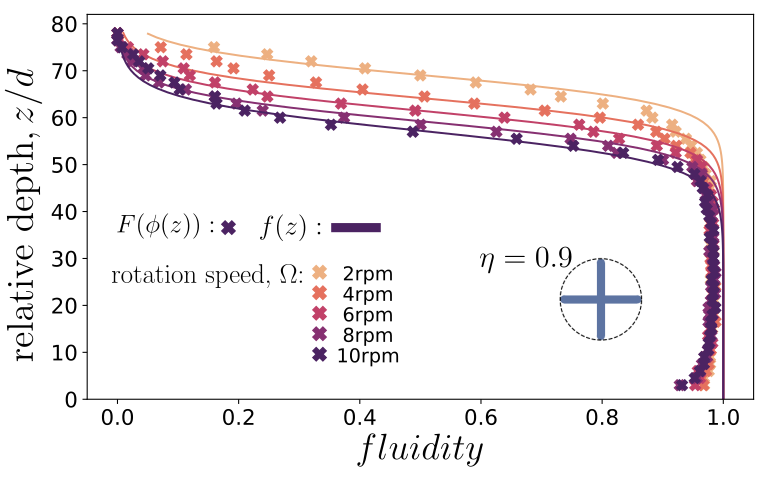}
  \put(-1,61){\bf{(b)}}
  \end{overpic}
  \end{minipage}
\caption{
Fluidity profile $f(z)$ (plain line) and the specific form of fluidity profile as a function of the packing fraction, $F(\phi(z))$ (scatter plot) for $\eta=0$ (a) and $\eta=0.9$ (b).}
\label{Fig_fluidity_fF}
\end{figure}

Figure \ref{Fig_fluidity_fF} show the evolution of both $f(z)$ (plain lines) and $F(\phi(z))$ (symbols) for $\eta=0$ (a) and $\eta=0.9$~rpm (b). As we can see, the data always collapses when $\xi$ is chosen equal to $\xi=2$ and $\lambda$ given in Fig.\ref{fig:paras_simus}a.

\section{Rewriting Eq.\ref{Eq_tau2} without neglecting extra terms}
\label{A_Correction_stress_with_xx_yy}
\begin{figure}[H]
  \begin{minipage}[t]{0.9\linewidth}
  \begin{overpic}[width=1\textwidth]{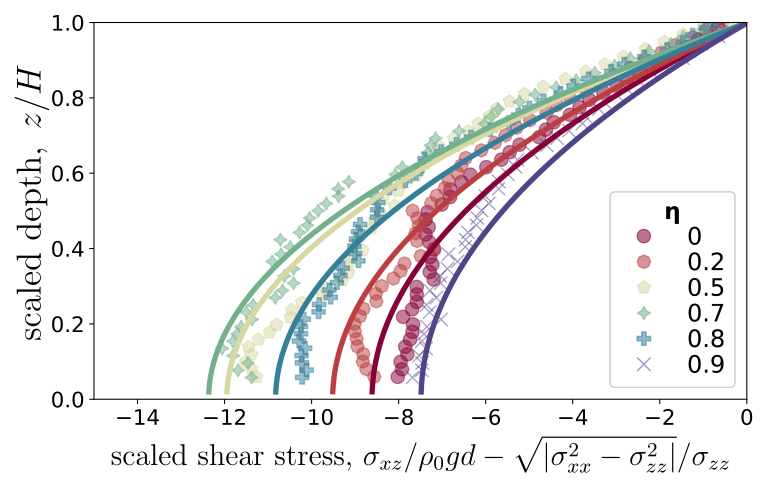}
  \put(-1,61){\bf{(a)}}
  \end{overpic}
  \end{minipage}
  
  \vspace{0.5cm}
  \begin{minipage}[t]{0.9\linewidth}
  \begin{overpic}[width=1\textwidth]{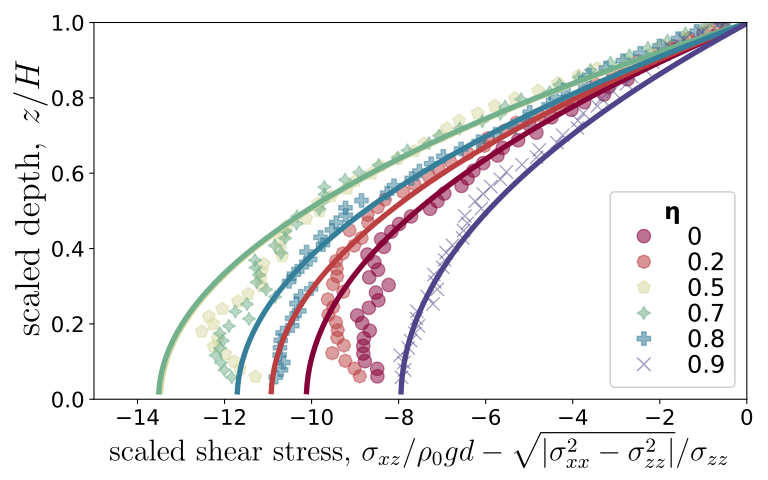}
  \put(-1,61){\bf{(b)}}
  \end{overpic}
  \end{minipage}
\caption{
Corrected shear stress, for different values of $\eta$ and for $\Omega=2$~rpm (a) and for different values of $\eta$ and for $\Omega=10$~rpm (b).}
\label{Fig_corrected_shear_stress}
\end{figure}

When developing Eq.\ref{Eq_tau2} we assumed that the stresses $\sigma_{xx}$ and $\sigma_{zz}$ were equal. This assumption is well verified in the bulk but may be too strong near the walls, more precisely at the bottom of the drum. Although the approximation proposed by Eq. \ref{Eq_tau2} reproduces the shear stress profiles very well, for all shapes and drum angular speed, it can be significantly improved without the previous assumption. In this case, the shear stress model reads as:
\begin{equation}
      \frac{\tau(z)}{\rho_0 gd}= \phi(z) \frac{H^2-z^2}{2Hd}\sin  \langle \theta \rangle + \frac{\sqrt{|\sigma_{xx}^2(z)-\sigma_{zz}^2(z)|}}{\sigma_{zz}(z)}.
\label{Eq:tau2_plus}
\end{equation}
Figure \ref{Fig_corrected_shear_stress} shows the evolution of the normalized shear stress profile $\tau$ for all shapes at $\Omega=2$~rpm (a) and $\Omega=10$~rpm (b). It also shows the approximation proposed by Eq.\ref{Eq:tau2_plus}, where $\sigma_{xx}(z)$ and $\sigma_{zz}(z)$ are measured directly from the simulations. As we can see, the approximation is more accurate than that given by Eq. \ref{Eq_tau2} in the region close to the drum border.

\section{Voronoi tessellation}
\label{Appendice_voronoi}
\begin{figure}[H]
  \centering
  \begin{minipage}[t]{0.45\linewidth}
  \includegraphics[width=0.95\textwidth, trim=2cm 4cm 8cm 0cm, clip]{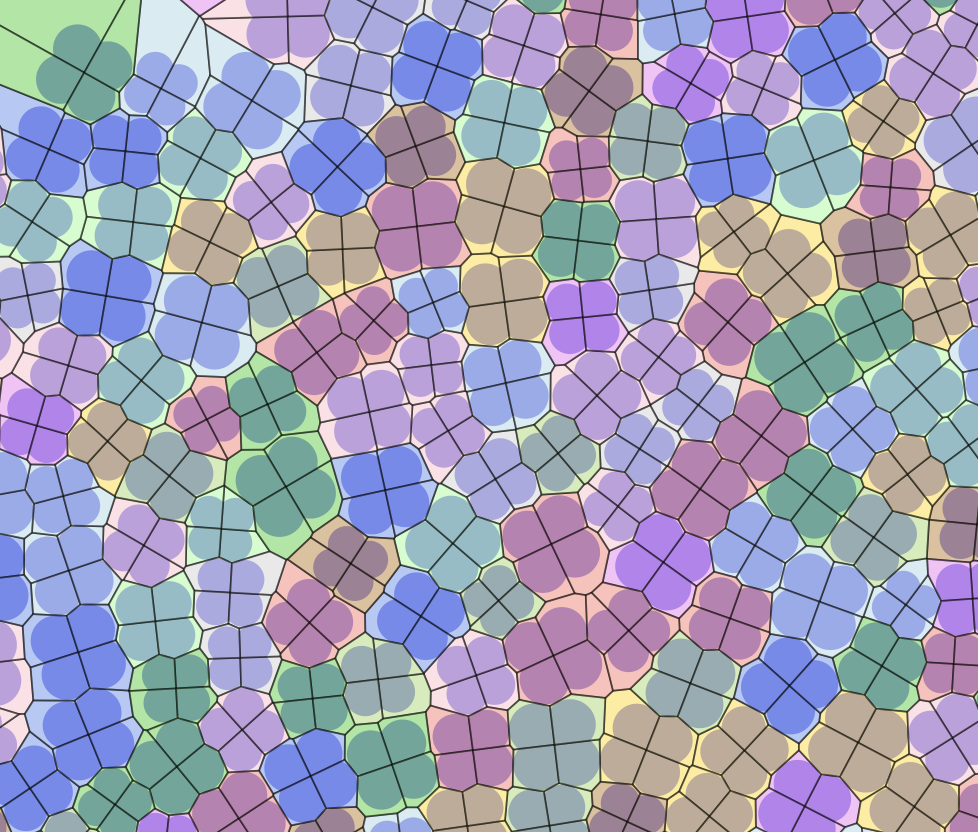}(a)
  \end{minipage}
  \hspace{0.2cm}
  \begin{minipage}[t]{0.45\linewidth}
  \includegraphics[width=0.95\textwidth, trim=6cm 8cm 16cm 3.7cm, clip]{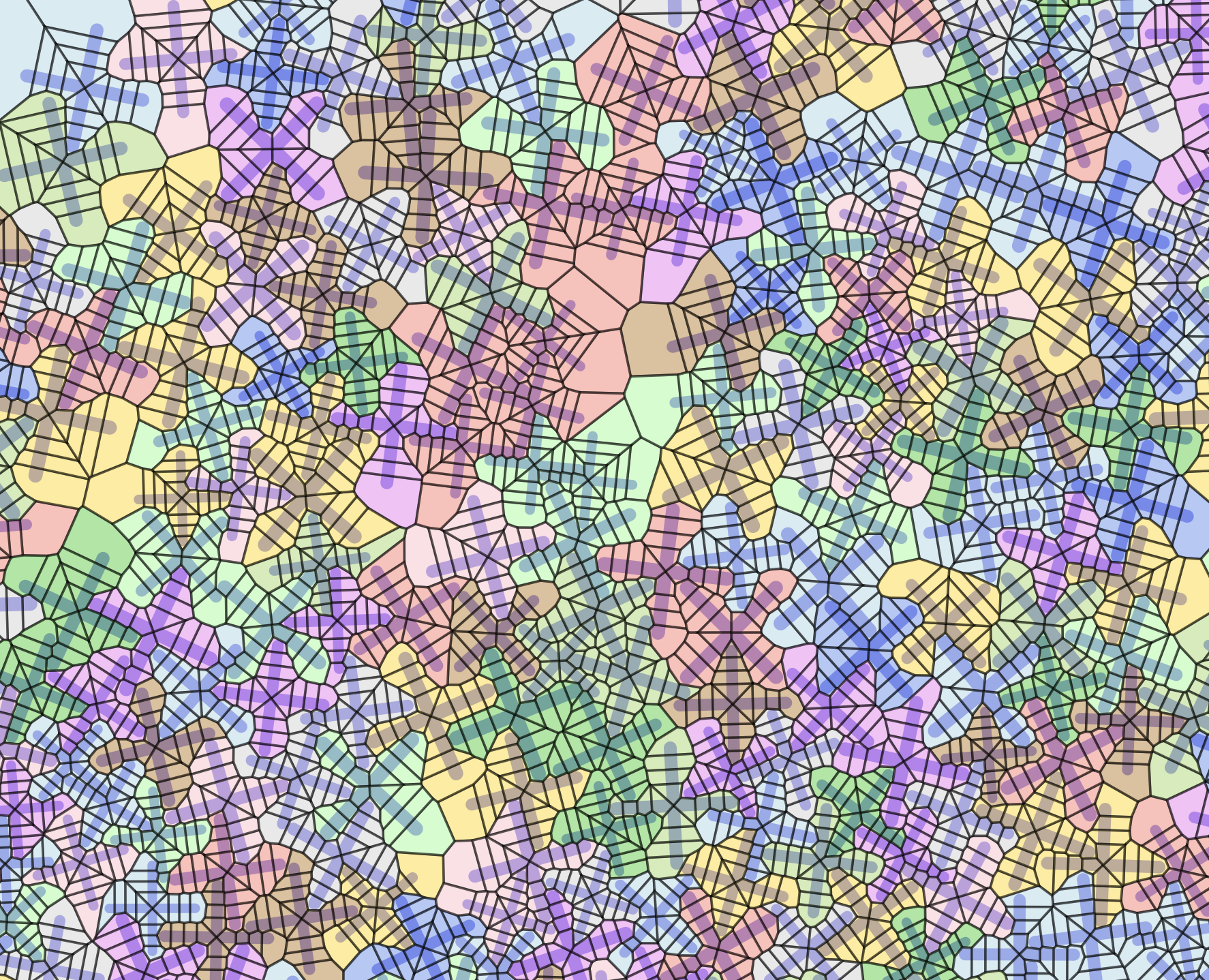}(b)
  \end{minipage}
\caption{Voronoi tessellation in a snapshot of simulation. (a): For concave particles with $\eta=0.5$. (b): For highly concave particles with $\eta=0.9$ }
\label{A_voronoi}
\end{figure}

To compute the packing fraction and stress, we used Voronoi tessellation to partition and determine the packing volume of each particle. Figure \ref{A_voronoi} shows a zoom illustrating the Voronoi meshing for highly non-convex grains.

\bibliography{References.bib}

\begin{thebibliography}{106}
\bibitem [{par()}]{particle_making}%
  \BibitemOpen
  \href@noop {} {}\bibinfo {note} {The mold has been made by cfo-outillage.fr.
  Injections have been done in IMT Mines Alès on a Krauss Maffei KM50/180CX.
  Batches of particles can be available upon request to the
  authors.}\BibitemShut {Stop}%
\end{thebibliography}

\end{document}